\documentclass[a4paper,12pt,twoside]{article}

\usepackage{amsfonts}
\usepackage{amsmath}
\usepackage{amsopn}
\usepackage{amsthm}
\usepackage{amssymb}
\usepackage{graphicx}
\usepackage{rotating}
\usepackage[]{natbib}
\usepackage[english]{babel}
\usepackage[utf8]{inputenc}
\usepackage[colorlinks=true,citecolor=blue,linkcolor=blue]{hyperref}
\usepackage{xspace}
\usepackage{float}
\usepackage{setspace}
\usepackage{dsfont}
\usepackage{booktabs}
\usepackage[T1]{fontenc}
\usepackage{centernot}
\usepackage{anysize}
\usepackage[labelfont=bf,labelsep=period]{caption}
\floatstyle{plaintop}
\restylefloat{table}
\marginsize{1.87cm}{1.87cm}{1.87cm}{1.87cm} 
\usepackage{fancyhdr}
\usepackage{algorithm}
\usepackage[noend]{algpseudocode}

\RequirePackage{txfonts}
\usepackage{times}

\newtheorem{Proposition}{Proposition}

\newtheorem{Theorem}{Theorem}

\renewcommand{\d}{\mathrm{d}}
\DeclareMathOperator{\re}{\mathbb{R}}

\newcommand{\abs}[1]{\left|#1\right|}

\newcommand{\ee}{\mathrm{e}}
\newcommand{\Prob}{\mathbb{P}}

\newcommand{\za}[1]{\stackrel{a}{#1}}
\newcommand{\zb}[1]{\stackrel{b}{#1}}
\newcommand{\zc}[1]{\stackrel{c}{#1}}

\allowdisplaybreaks

\begin{document}

\def\figureautorefname{Figure}
\def\algorithmautorefname{Algorithm}
\def\sectionautorefname{Section}
\def\subsectionautorefname{Section}
\def\Propositionautorefname{Proposition}
\def\Theoremautorefname{Theorem}
\def\Lemmaautorefname{Lemma}
\def\Assumptionautorefname{Assumption}
\def\Corollaryautorefname{Corollary}
\renewcommand*\footnoterule{}

\title{Robustness against conflicting prior information in regression}

\author{Philippe Gagnon$^{1}$}

\maketitle

\thispagestyle{empty}

\noindent $^{1}$Department of Mathematics and Statistics, Universit\'{e} de Montr\'{e}al, Canada.

\begin{abstract}

    Including prior information about model parameters is a fundamental step of any Bayesian statistical analysis. It is viewed positively by some as it allows, among others, to quantitatively incorporate expert opinion about model parameters. It is viewed negatively by others because it sets the stage for subjectivity in statistical analysis. Certainly, it creates problems when the inference is skewed due to a conflict with the data collected. According to the theory of conflict resolution \citep{o2012bayesian}, a solution to such problems is to diminish the impact of conflicting prior information, yielding inference consistent with the data. This is typically achieved by using heavy-tailed priors. We study both theoretically and numerically the efficacy of such a solution in a regression framework where the prior information about the coefficients takes the form of a product of density functions with known location and scale parameters. We study functions with regularly-varying tails (Student distributions), log-regularly-varying tails (as introduced in \cite{desgagne2015robustness}), and propose functions with slower tail decays that allow to resolve any conflict that can happen under that regression framework, contrarily to the two previous types of functions. The code to reproduce all numerical experiments is available online.\footnote{See ancillary files on \href{https://arxiv.org/abs/2110.09556}{arXiv:2110.09556}.}

\end{abstract}


\section{Introduction}\label{sec_intro}

\subsection{Context}\label{sec:context}

In Bayesian analysis, prior information about the parameters of a regression model is included using prior distributions. Consider a model $Y \sim \Prob_{\eta, \boldsymbol\psi}$, with $\eta := \mathbf{x}^T \boldsymbol\beta$ being a linear predictor. For this regression model, the parameters are $\boldsymbol\beta$ and $\boldsymbol\psi$, where $\boldsymbol\beta := (\beta_1, \ldots, \beta_p)^T \in \re^p$ are the regression coefficients, with $p$ a positive integer, and $\boldsymbol\psi$ is a vector formed of, e.g., scale or shape parameters; $\mathbf{x}$ is a known vector of covariates. This regression framework encompasses linear regression, generalized linear models (GLMs) and generalized additive models (when estimated using a spline representation). In this paper, we study the impact on statistical inference of prior information in conflict with the data collected for different types of prior distributions. Our study rests heavily on the form of the prior distributions which will be seen to be a product form where each regression-coefficient density has known location and scale parameters, justifying the introduction of such a study within a regression framework. We focus on situations where the prior information that is in conflict is about regression coefficients, the latter being typically of main interest which makes them more likely to be assigned informative prior distributions. Additionally, we focus on situations where the prior distributions on the coefficients are used to include prior information about the latter, not to regularize the model (contrarily to in, e.g., \cite{johnstone2004needles, park2008bayesian, carvalho2010horseshoe}), even though the study conducted here may be helpful to develop regularization strategies. Furthermore, we focus on linear regression as a special case of the general regression framework described above. This will allow to state precise theoretical results about the behaviour of the posterior distribution in conflicting situations, depending on the type of prior distributions employed. We will explain why and how the results presented apply in the general regression framework.

From now on, we thus consider that $Y = \mathbf{x}^T \boldsymbol\beta + \sigma \varepsilon$ with $\varepsilon \sim f$, which is equivalent to $Y \sim (1 / \sigma) f((\, \cdot \, - \mathbf{x}^T \boldsymbol\beta) / \sigma)$, where $\varepsilon$ is a standardized error term, $\sigma > 0$ is a scale parameter and $f$ is a distribution; to simplify, $f$ is also used to denote the probability density function (PDF) associated to the distribution. When all covariates are continuous (i.e.\ when they all take values in uncountable totally-ordered sets), it is recommended to define the prior distribution of the regression coefficients using a conditional-independence structure (see, e.g., \cite{1984west431} and \cite{raftery1997bayesian}):
\begin{align}\label{eqn:prior}
 \pi(\boldsymbol\beta \mid \boldsymbol\psi) = \pi(\boldsymbol\beta \mid \sigma) := \prod_{j = 1}^p \pi_j(\beta_j \mid \sigma) := \prod_{j = 1}^p \frac{\lambda_j}{\sigma} g_j\left(\frac{\lambda_j}{\sigma} (\beta_j - \mu_j) \right),
\end{align}
where all $g_j$ are strictly positive bounded density functions that are symmetric with respect to 0, and $\mu_j \in \re$ and $\sigma / \lambda_j > 0$ play the role of location and scale parameters, respectively; $\mu_j$ and $\lambda_j$ are considered to be known and chosen by the user. In the following, we consider to simplify that all covariates are continuous; the theoretical results hold even when this is not the case, but under more technical assumptions. Note that, to simplify the notation, $g_j$ is also used to denote the distribution associated to the density.

Determining the outcome of conflicting prior information under a general structure of dependence in between the coefficients requires a multivariate analysis and depends strongly on the structure of dependence. The conditional-independence structure presented above allows to simplify the problem and transform the multivariate analysis into several univariate analyses, in addition to enabling the exploitation of existing conflict-resolution techniques that are based on univariate heavy-tailed distributions (and for which the relevant literature will be presented when describing the techniques below). A general and multivariate analysis to determine the outcome of conflicting prior information is beyond the scope of this manuscript.

\subsection{Conflicts}

\begin{figure}[ht]
\centering
$\begin{array}{cc}
    \vspace{-2mm}\hspace{-2mm} \includegraphics[width=0.50\textwidth]{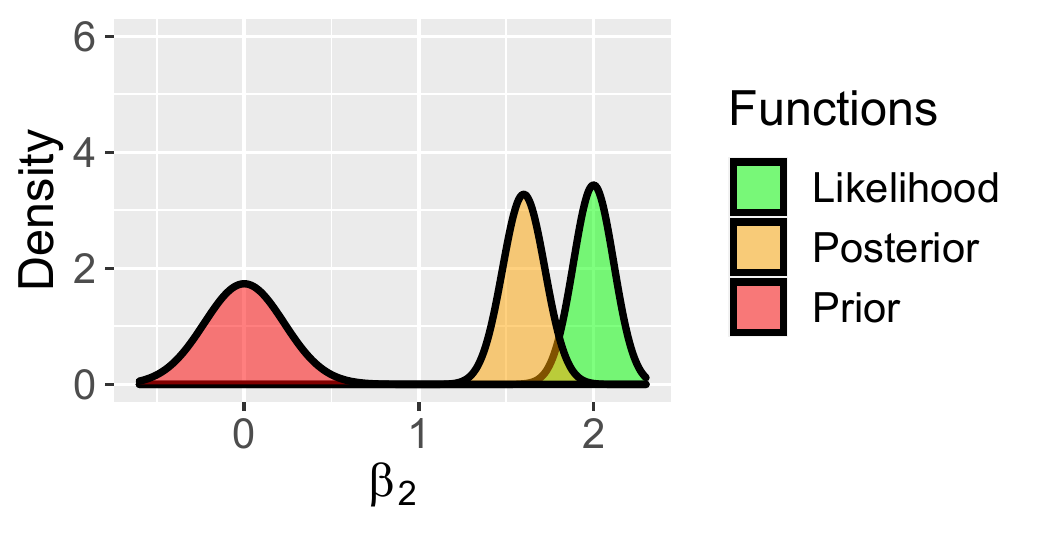} & \hspace{-3mm} \includegraphics[width=0.50\textwidth]{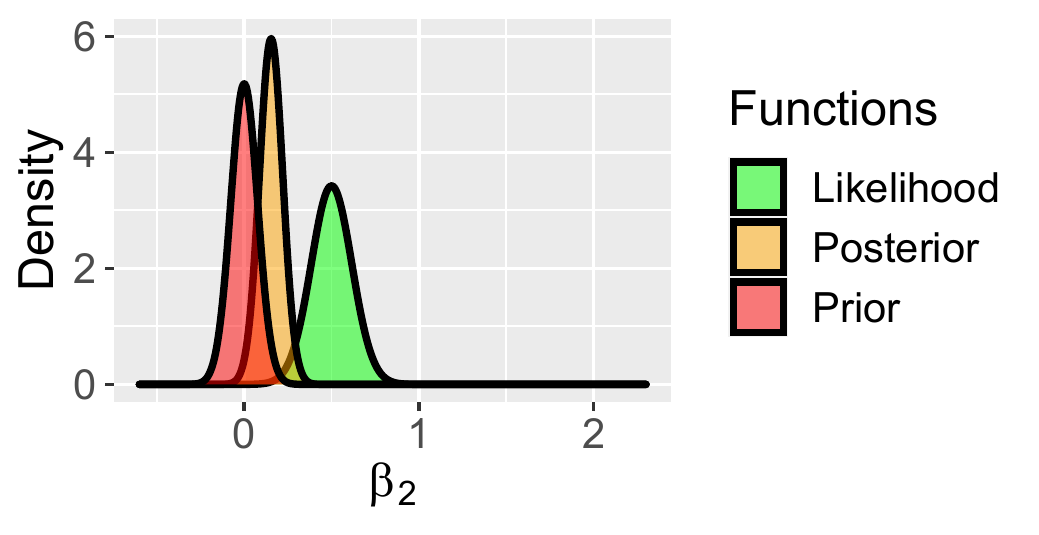} \cr
    \hspace{-20mm} \textbf{(a)} & \hspace{-20mm} \textbf{(b)}
\end{array}$
  \vspace{-3mm}
\caption{\small Two examples of conflicts where $y = 0 + \beta_2 x_{2} + \sigma\varepsilon$, $\varepsilon \sim \mathcal{N}(0, 1)$, $g_2 = \mathcal{N}(0, 1)$, the sample size is $n = 100$, the prior on $\sigma$ is an inverse-gamma with shape and scale parameters of $n / 2$ each, the variables are standardized, and: (a) $\mu_2 = 0$, $\lambda_2 = \sqrt{n} / 2$ and $\hat{\beta}_2^{\text{OLS}} = 2$, (b) $\mu_2 = 0$, $\lambda_2 = 1.5 \sqrt{n}$ and $\hat{\beta}_2^{\text{OLS}} = 0.5$; $\hat{\beta}_2^{\text{OLS}}$ is the ordinary-least-squares (OLS) estimate which corresponds to the maximum likelihood estimate in that case; in this figure, the likelihood function is normalized to make it a PDF}\label{fig:conflicts}
\end{figure}

\begin{figure}[ht]
\centering
$\begin{array}{cc}
    \vspace{-2mm}\hspace{-2mm} \includegraphics[width=0.50\textwidth]{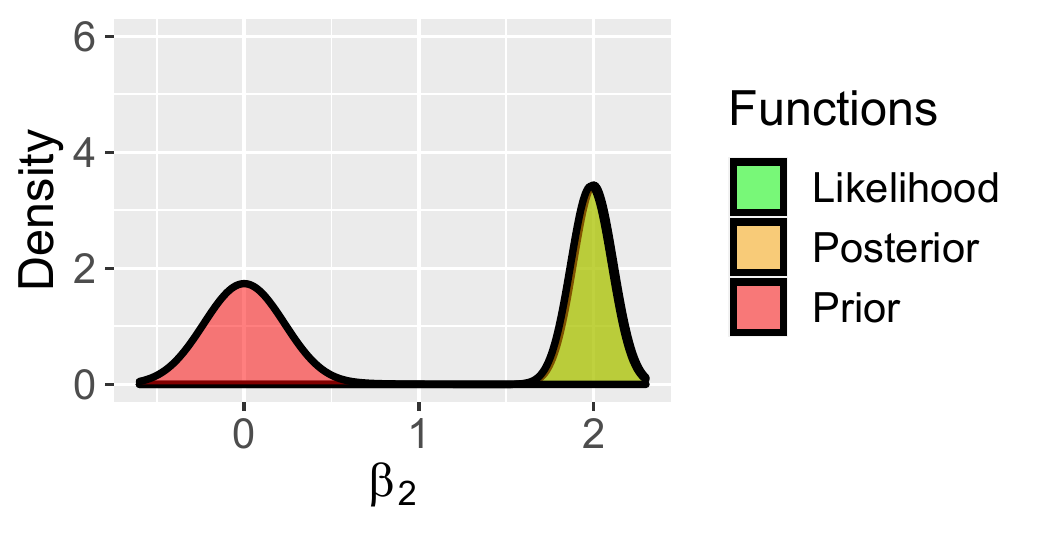} & \hspace{-3mm} \includegraphics[width=0.50\textwidth]{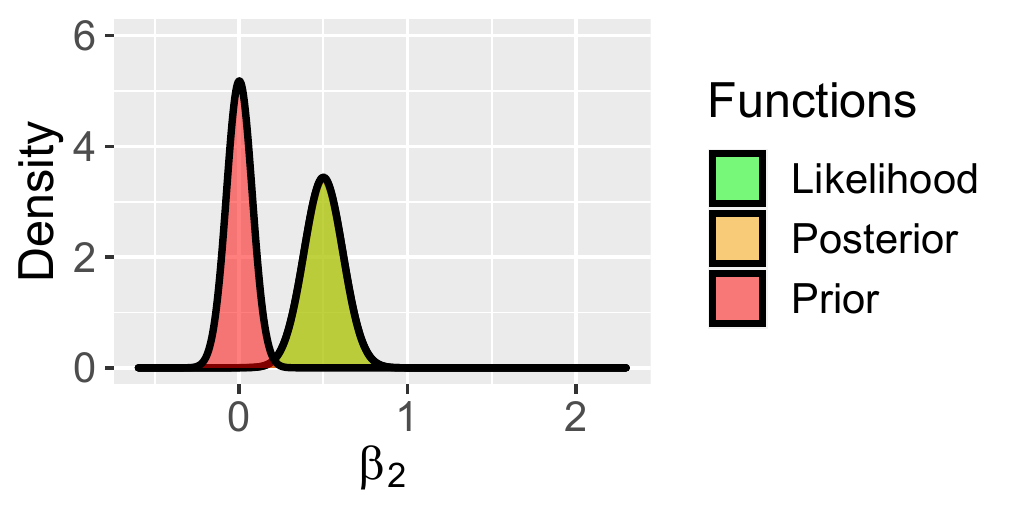} \cr
    \hspace{-20mm} \textbf{(a)} & \hspace{-20mm} \textbf{(b)}
\end{array}$
  \vspace{-3mm}
\caption{\small Same two examples of conflicts as in \autoref{fig:conflicts} with the only difference being that in: (a) $g_2 = \text{LPTN}$ as defined in \autoref{sec:heavy_priors} with $\rho = 0.95$; (b) $g_2 = \text{CTN}$ as defined in \autoref{sec:heavy_priors} with $\varrho = 0.98$}\label{fig:resolution}
\end{figure}

In normal linear regression, where $f = \mathcal{N}(0, 1)$, conjugate priors are often employed, i.e.\ $g_j = \mathcal{N}(0, 1)$ in \eqref{eqn:prior} and $\sigma^2$ follows an inverse-gamma distribution. A prior is in conflict with the likelihood when the areas where these function have high densities are significantly different (\autoref{fig:conflicts}). When both the prior and the likelihood are normal (given $\sigma$), an undesirable compromise follows: the posterior concentrates its mass on an area in between those with high prior and likelihood densities. This is a consequence of the slimness of the normal tails: the area where the likelihood function has high density is in the tails of the prior density which have an exponential decay, penalizing extremely for such parameter values, and the same holds if we inverse the role of the prior and likelihood in the previous statement. The areas with high prior and likelihood densities thus become \textit{a posteriori} less probable than an area in between, representing how a conflict is dealt with by that Bayesian modelling and an ineffective way of resolving a conflict. Indeed, the posterior distribution is not consistent with either of the sources of information. Here we consider that the data model is well specified and that the data can be trusted; the information about the parameters carried by the data is thus favoured to the prior information when they conflict. Therefore, we consider that a conflict is (effectively) resolved when the conflicting prior information is discarded so as to yield a posterior distribution consistent with the data (\autoref{fig:resolution}).

We acknowledge that the assumption of a well specified data model and that the data can be trusted is strong, but we make this assumption in order to be able to focus on robustness against conflicting prior information. We can, for instance, allow for some sort of misspecification and a potential presence of extreme/erroneous data on top of conflicting prior information by considering that the data set may contain outliers, and obtain similar theoretical results as those presented in the next sections. This is because we allow for the regression model to have an heavy-tailed error distribution. It is however beyond the scope of this manuscript to analyse the situation of potential presence of outliers and present related results.

\normalsize

\normalsize

Plots (a) and (b) in Figures \ref{fig:conflicts} and \ref{fig:resolution} are meant to represent two distinct conflicting situations: (a) one where the conflict is due to a prior location that is significantly different than that of the likelihood, and (b) one where it is due to a extremely small prior scaling. We analyse both situations theoretically and numerically in the next sections. The theoretical analysis will be conducted under an asymptotic regime. In the first situation, the asymptotic regime corresponds to one where the distance between the red and green areas in \autoref{fig:conflicts} (a) increases without bounds, which is mathematically modelled by the red one moving away, i.e. $\mu_j \rightarrow \pm \infty$; in the second situation, we will consider that $\lambda_j \rightarrow \infty$. It will be seen that a prior distribution which does lead to a resolution of conflict in the first situation does not necessarily in the second one. The first situation can be think of as one where a practitioner was wrong about the parameter location, but incorporated a moderate confidence by using a moderate prior scaling. In the second situation, in addition to being wrong about the parameter location (but less severely than in the first situation), the practitioner was also overly confident; this conflicting situation could have been avoided by using a less concentrated prior. The latter is also true in the first situation, but the prior would need to be much less concentrated. That is why in one case we consider that the problematic aspect is the location, whereas we consider that it is the scaling in the other one.

A natural way to achieve effective conflict resolution is to have recourse to heavy-tailed distributions; in situations like those presented in \autoref{fig:resolution}, the areas where the likelihood functions have high densities are still in the tails of the prior densities, but more weight is assigned to those tails, thus penalizing less for such extreme situations. This strategy dates back to \cite{de1961bayesian} with a first analysis in \cite{lindley1968choice}, followed by an introduction of a formal theory in \cite{dawid1973posterior}, \cite{hill1974coherence} and \cite{o1979outlier}. For a recent review of Bayesian heavy-tailed models and conflict resolution, see \cite{o2012bayesian}. In the latter paper, it is noted that there exists a gap between the models formally covered by the theory of conflict resolution and models commonly used in practice. The latest developments focus on situations where the conflicting information is carried by outlying data points in location-scale models \citep{desgagne2015robustness} and linear regression \citep{DesGag2019, gagnon2020, gagnon2020PCR, hamura2020log, gagnon2022theoretical}. The present paper contributes to the expansion of the theory of conflict resolution by covering conflicting prior information in regression.

We consider that in the ideal situation where it is guaranteed that the priors will not conflict with the data that will be collected, prior information about the regression coefficients is included by setting $g_j = \mathcal{N}(0, 1)$, which is the favoured choice in practice. Given that we consider that the data model is well specified and that the data can be trusted, the distributions that we alter to achieve effective conflict resolution are thus the $g_j$'s. A desideratum of the resulting heavy-tailed priors is to yield similar inference to the informative light-tailed priors they replace in the absence of conflict. In the following, we study three alternatives to the normal distribution with three different types of tail decays: a first one with regularly-varying tails, a second one with log-regularly-varying tails \citep{desgagne2015robustness}, and a third one with constant tails. They are all presented in \autoref{sec:heavy_priors} in which an overview of their advantages and disadvantages is also provided. In \autoref{sec:results}, their efficacy are precisely characterized through theoretical results. An extensive simulation study is next provided in \autoref{sec:simulation} to show how these theoretical results translate in practice. The manuscript finishes in \autoref{sec:conclusion} with retrospective comments. All proofs of theoretical results are deferred to Appendix A (supplementary material). Some details of the simulation study are presented in Appendix B (supplementary material).

\section{Heavy-tailed priors}\label{sec:heavy_priors}

We start in \autoref{sec:Student} by presenting the main characteristics of the most commonly employed alternative to the normal distribution in conflict resolution, the Student distribution. Even in the least problematic situation, which is that where the conflict is due to a prior location that is significantly different than that of the likelihood, it will be seen to partially resolve conflicts. We next provide a description of the \textit{log-Pareto-tailed normal} (LPTN) distribution in \autoref{sec:LPTN}, which has the ability to wholly discard the prior information in that situation. This distribution was introduced by \cite{desgagne2015robustness}. Its density exactly matches that of the standard normal on the interval $[-\tau, \tau]$, where $\Prob(- \tau \leq \mathcal{N}(0, 1) \leq \tau) = \rho$. Outside of this area, the tails of this continuous density are log-regularly varying \citep{desgagne2015robustness}, and behave as log-Pareto tails, i.e.\ $(1 / |z|)(1 / \log |z|)^\theta$, hence its name. The only free parameter of this distribution is $\rho$: the parameter $\theta$ is a function of $\rho$ and $\tau$, the latter being itself a function of $\rho$. Even with such heavy tails, the LPTN distribution leads to an ineffective conflict resolution when the conflict is due to a small prior scaling. In response to this problem, we introduce in \autoref{sec:CTN} the \textit{constant-tailed normal} (CTN) distribution, which, like the LPTN distribution, has a density that matches that of the standard normal on a central interval, but with constant tails.

\subsection{Student distribution}\label{sec:Student}

\begin{figure}[ht]
\centering
$\begin{array}{cc}
    \vspace{-2mm}\hspace{-2mm} \includegraphics[width=0.50\textwidth]{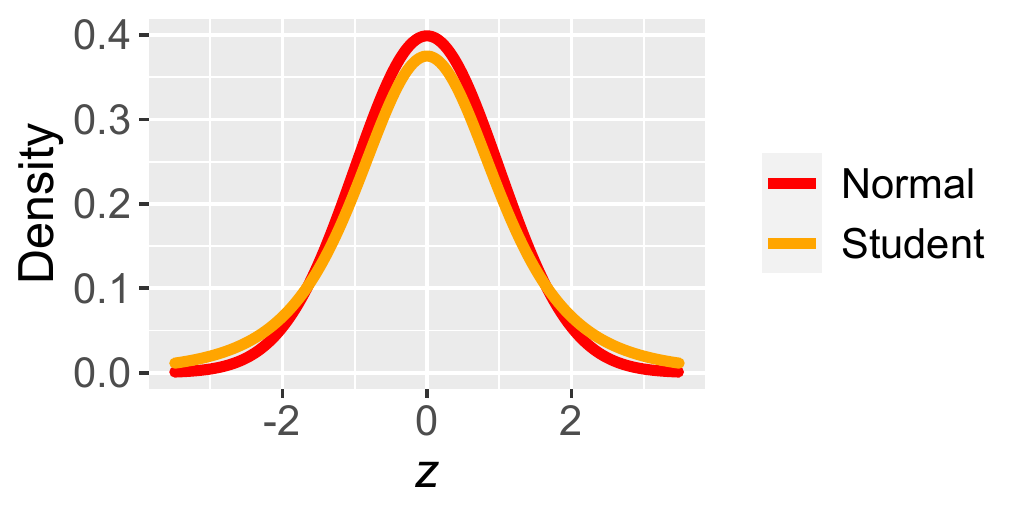} & \hspace{-3mm} \includegraphics[width=0.50\textwidth]{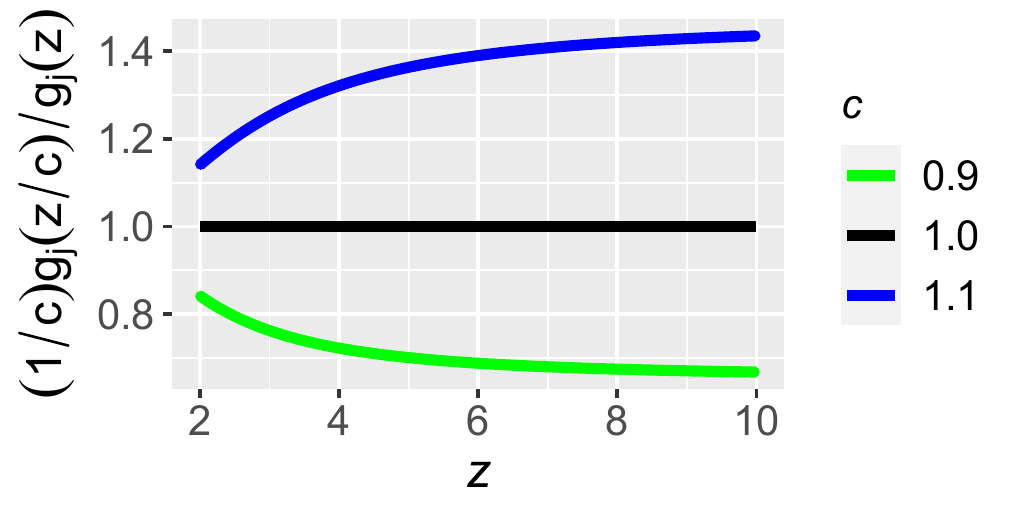} \cr
    \hspace{-13mm} \textbf{(a)} & \hspace{-7mm} \textbf{(b)}
\end{array}$
\caption{\small (a) PDFs of the standard normal distribution and the Student distribution with 4 degrees of freedom, (b) ratios of two Student PDFs with 4 degrees of freedom where the PDF at the numerator as an additional scale parameter of value $c$}\label{fig:Student}
\end{figure}

The Student distribution is without a doubt the favourite heavy-tailed alternative to the standard normal distribution. A reason for this is because its density shares important characteristics with the standard normal one, like a bell shape and symmetry around 0. We show this in \autoref{fig:Student} (a) for a Student distribution with 4 degrees of freedom, which represents a good compromise between heavy tails and close similarity with the normal distribution. In \autoref{fig:Student} (b), we show how the ratio $(1 / c)g_j(z / c) / g_j(z)$ behaves as $z \rightarrow \infty$ when $g_j$ is the PDF of a Student distribution with 4 degrees of freedom to graphically illustrate its regularly-varying property, a property that is discussed in greater detail below.

\normalsize

Employing Student prior distributions instead of normal ones for conflict resolution in regression has been explored before; see, e.g., \cite{1984west431} and \cite{mutlu68robust}. However, the focus of previous papers was different than that of the current one, which is to compare that alternative to normal prior distributions to other alternatives through an extensive theoretical and numerical analysis. In \cite{1984west431}, for instance, the focus is rather to study the use of the Student distribution in a context of robustness against outliers; the Student is viewed as a member of a specific family of alternatives to normal distributions, that of scale mixtures of normal distributions.

The tails of the Student density are regularly varying, implying that for any fixed $\lambda_j, \sigma, \beta_j$,
\begin{align}\label{eqn:limit_stu}
 \lim_{\mu_j \rightarrow \pm \infty} \frac{\frac{\lambda_j}{\sigma} g_j\left(\frac{\lambda_j}{\sigma} (\beta_j - \mu_j) \right)}{g_j(\mu_j)} = \lim_{\mu_j \rightarrow \pm \infty} \frac{\lambda_j}{\sigma} \left(\frac{\gamma + \mu_j^2}{\gamma + (\lambda_j / \sigma)^2 (\beta_j - \mu_j)^2}\right)^{\frac{\gamma + 1}{2}} = \left(\frac{\sigma}{\lambda_j}\right)^\gamma,
\end{align}
where $\gamma$ is the degrees of freedom. Examining the limiting behaviour of prior densities is an important step in understanding the limiting behaviour of the posterior distribution in conflicting situations. Indeed, given that the posterior density is the normalized product of the prior densities and the likelihood function, the limit above suggests that a conflicting prior density (due to a significantly different location) behaves in the limiting posterior distribution like $(\sigma / \lambda_j)^\gamma g_j(\mu_j) \propto \sigma^\gamma$. The theoretical results in \autoref{sec:results} precisely characterize the behaviour of the limiting posterior distribution, depending on the conflicting situation and the priors employed. With a Student prior distribution, conflicting information is partially rejected as a trace remains, $\sigma^\gamma$. Ideally, conflicting information is wholly rejected as its source becomes increasingly remote \citep{1984west431}, which translates into a prior density which behaves asymptotically like $g_j(\mu_j) \propto 1$. This explains why we say that the Student distribution only partially resolves conflicts due to significantly different locations. The existence of that trace is a consequence of employing a prior density with insufficiently heavy tails. Indeed, it will be seen in \autoref{sec:LPTN} that the limit of the ratio in \eqref{eqn:limit_stu} when instead setting $g_j$ to a LPTN distribution is 1. The trace has an impact on the limiting posterior variability of all coefficients which is seen to be more or less significant depending on the degrees of freedom, the sample size and the number of conflicting prior densities (this is shown explicitly in \autoref{sec:simulation}). When the sample size is large relatively to the degrees of freedom and the number of conflicting prior densities, as in the numerical experiment of \autoref{sec:simulation}, the impact is small. The insufficiently heavy tails however make the convergence to the limiting posterior distribution slower, comparatively with other alternatives with heavier tails, implying a slower partial resolution of conflicts. We finish this section by noting that the Student prior density converges to a point mass at $\mu_j$ when $\lambda_j \rightarrow \infty$, which makes it ineffective at resolving conflicts due to extremely small prior scalings.

\subsection{LPTN distribution}\label{sec:LPTN}

\begin{figure}[ht]
 \centering
$\begin{array}{cc}
    \vspace{-2mm}\hspace{-2mm} \includegraphics[width=0.50\textwidth]{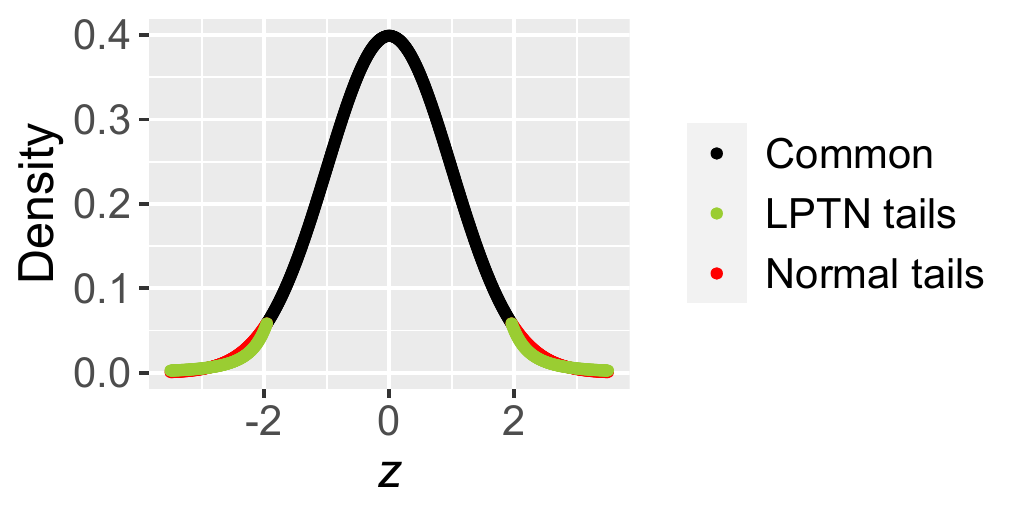} & \hspace{-3mm} \includegraphics[width=0.50\textwidth]{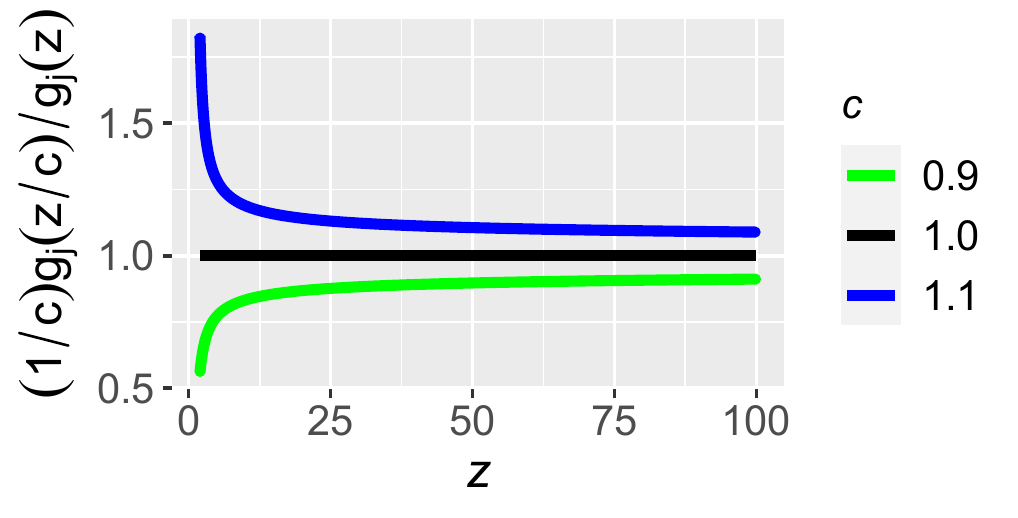} \cr
    \hspace{-13mm} \textbf{(a)} & \hspace{-7mm} \textbf{(b)}
\end{array}$
\caption{\small (a) PDFs of the standard normal distribution and the LPTN distribution with $\rho = 0.95$, (b) ratios of two LPTN PDFs with $\rho = 0.95$ where the PDF at the numerator as an additional scale parameter of value $c$}\label{fig:LPTN}
\end{figure}

The density of the LPTN distribution is as follows:
  \begin{align*}
  g_{\text{LPTN}}(z) := \left\{
                                                    \begin{array}{lcc}
                                                      \varphi(z)  & \text{ if } & \abs{z}\leq \tau, \\
                                                      \varphi(\tau)\,\frac{\tau}{|z|}\left(\frac{\log \tau}{\log |z|}\right)^{\theta} & \text{ if } & \abs{z}>\tau, \\
                                                    \end{array}
\right.
  \end{align*}
 where $z\in\re$, and $\tau>1$ and $\theta>1$ are functions of a parameter $\rho \in (2\Phi(1) - 1, 1) \approx (0.6827, 1)$ with
  \begin{align*}
 & \tau=\Phi^{-1}((1+\rho)/2) := \{\tau : \Prob(-\tau \leq Z \leq \tau)= \rho \,\text{ for }\, Z\, \sim \, \mathcal{N}(0,1)\}, \\
 & \theta=2(1-\rho)^{-1}\varphi(\tau) \, \tau \log(\tau) + 1,
 \end{align*}
  $\varphi$, $\Phi$ and $\Phi^{-1}$ being the PDF, cumulative distribution function (CDF) and inverse CDF of a standard normal, respectively. A LPTN density with $\rho = 0.95$ is presented in \autoref{fig:LPTN} (a). This choice of value for $\rho$ yields, like the Student with 4 degrees of freedom in \autoref{fig:Student} (a), a good compromise between heavy tails and close similarity with the normal distribution. In \autoref{fig:LPTN} (b), we show how the ratio $(1 / c)g_j(z / c) / g_j(z)$ behaves as $z \rightarrow \infty$ when $g_j$ is the PDF of a LPTN distribution with $\rho = 0.95$ to graphically illustrate its log-regularly-varying property, a property that is discussed in greater detail below.

  As was seen in \autoref{fig:resolution} (a), this slightly modified version of the normal distribution with $\rho = 0.95$ can resolve conflicts very effectively; the likelihood function and posterior density are indeed on top of each other in that figure. The parameter $\rho$, chosen by the user, represents the mass of the central part that exactly matches the $\mathcal{N}(0, 1)$ density. The value $0.95$ has been seen to be a good choice for robustness against outliers in linear regression (see \cite{gagnon2020} and \cite{gagnon2020PCR}). We analyse the impact of the value of $\rho$ in the present context in \autoref{sec:simulation}.

\normalsize

 An advantage of Student distributions over LPTN distributions is that their densities are smooth. Indeed, we see in \autoref{fig:LPTN} (a) that in order to obtain a density that exactly matches that of the standard normal on an interval, while having heavier tails and being continuous, the LPTN density has to decrease quicker than the normal one for a short interval beyond $|z| = \tau$, making the derivative of the LPTN density discontinuous at $|z| = \tau$. The smoothness of a posterior density has an impact on the efficiency of the numerical methods used to approximate integrals with respect to the associated posterior distribution. With densities having discontinuous derivatives, one might wonder if it is even possible to apply numerical methods explicitly exploiting gradients of log posterior densities, like Metropolis-adjusted Langevin algorithms \citep{roberts1996exponential} and Hamiltonian Monte Carlo (HMC, \cite{Duane1987}), which both are Markov-chain Monte Carlo methods. Given that the discontinuity points of the LPTN derivative have null measure, these methods can be applied. HMC has in fact been employed to sample from the resulting posterior distributions to compute estimates and posterior variances in \autoref{sec:simulation} and no problems have been encountered.

 The main advantage of LPTN distributions over Student distributions is that the limit of the ratio of densities analogous to \eqref{eqn:limit_stu} is equal to 1, as established in the next proposition, showing their ability at effectively resolving conflicts due to significantly different locations.
  \begin{Proposition}[Asymptotic location--scale invariance]\label{prop:location-scale}
If $g_j = g_{\text{LPTN}}$, we have that for any fixed $\lambda_j, \sigma, \beta_j$,
\begin{align*}
 \lim_{\mu_j \rightarrow \pm \infty} \frac{\frac{\lambda_j}{\sigma} g_j\left(\frac{\lambda_j}{\sigma} (\beta_j - \mu_j) \right)}{g_j(\mu_j)} = \lim_{\mu_j \rightarrow \pm \infty} \frac{|\mu_j|}{|\beta_j - \mu_j|}\left(\frac{\log |\mu_j|}{\log (\lambda_j / \sigma) |\beta_j - \mu_j|}\right)^\theta = 1.
\end{align*}
\end{Proposition}
 The property of asymptotic location--scale invariance is shared by all log-regularly-varying distributions (LRVDs, \cite{desgagne2015robustness}). Most members of this family of distributions supported on the real line are distributions which tails were not originally log-Pareto ones (like the standard normal distribution), but their tails were replaced to reach the desired tail decay, i.e.\ $(1 / |z|)(1 / \log |z|)^\theta$ (like the LPTN distribution). The strategy of replacing the tails of a light-tailed prior distribution by heavy tails to attain an asymptotic location--scale invariance can thus be applied even when the light-tailed prior is not a normal. A distribution which is LRVD, but with originally log-Pareto tails, is the log transformation of a Pareto distribution. This distribution is not often employed because its density has a spike at zero, and is thus less appealing than smooth bell curves like normal and Student densities.

 Another advantage of LPTN distributions over Student distributions is that its density is even more similar to the normal one, yielding more similar inferences in the absence of conflict, as will be seen in \autoref{sec:simulation}. Although there are advantages in using LPTN prior distributions, there are disadvantages; one has been mentioned above, but the main disadvantage is that they do not allow, like all LRVDs, to resolve conflicts due to large $\lambda_j$. Indeed, for any fixed $\beta_j, \mu_j \in \re $ and $\sigma > 0$, with $\beta_j \neq \mu_j$, if $g_j = g_{\text{LPTN}}$ and $\lambda_j$ is large enough,
\begin{align}\label{eqn:LPTN_lambda}
 \frac{\lambda_j}{\sigma} g_j\left(\frac{\lambda_j}{\sigma} (\beta_j - \mu_j) \right) &= \varphi(\tau) \, \frac{\lambda_j}{\sigma}\frac{\tau}{(\lambda_j / \sigma) |\beta_j - \mu_j|}\left(\frac{\log \tau}{\log [(\lambda_j / \sigma) |\beta_j - \mu_j|]}\right)^\theta \cr
 &= \varphi(\tau) \, \frac{\tau}{|\beta_j - \mu_j|}\left(\frac{\log \tau}{\log \lambda_j}\right)^\theta \left(\frac{1}{1 + [\log |\beta_j - \mu_j| / \sigma] / \log \lambda_j}\right)^\theta,
\end{align}
which is asymptotically equivalent as $\lambda_j \rightarrow \infty$ to
\[
 \varphi(\tau) \, \frac{\tau}{|\beta_j - \mu_j|}\left(\frac{\log \tau}{\log \lambda_j}\right)^\theta \propto \frac{1}{|\beta_j - \mu_j|}.
\]
Analogously to Student priors in the previous section (but not for the same type of conflict), conflicting information is partially rejected as a trace remains, $|\beta_j - \mu_j|^{-1}$. The latter includes information about the location significantly differently than the normal distribution does. Additionally, the rightmost term in \eqref{eqn:LPTN_lambda} converges to 1 slowly (because the speed at which $[\log |\beta_j - \mu_j| / \sigma] / \log \lambda_j$ vanishes is dictated by that at which $\log \lambda_j$ goes to infinity). This implies that the conflicting information is slowly (in addition to partially) rejected, which will be observed empirically in \autoref{sec:simulation}. For all these reasons, we consider that LPTN prior distributions are ineffective at resolving conflicts due to small prior scalings, motivating the introduction of different heavy-tailed alternatives to normal prior distributions.

\subsection{CTN distribution}\label{sec:CTN}

\begin{figure}[ht]
   \centering
$\begin{array}{cc}
    \vspace{-2mm}\hspace{-2mm} \includegraphics[width=0.50\textwidth]{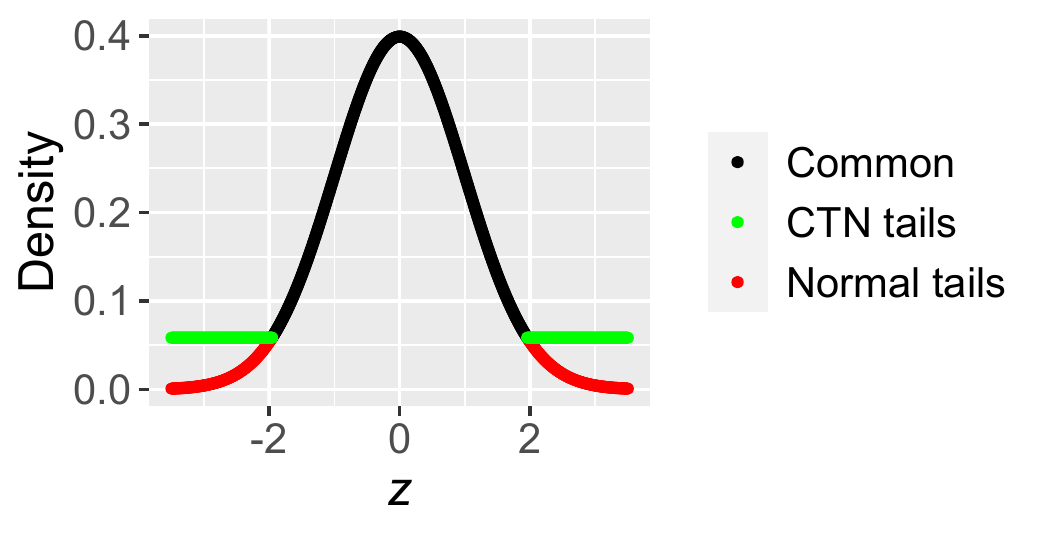} & \hspace{-3mm} \includegraphics[width=0.50\textwidth]{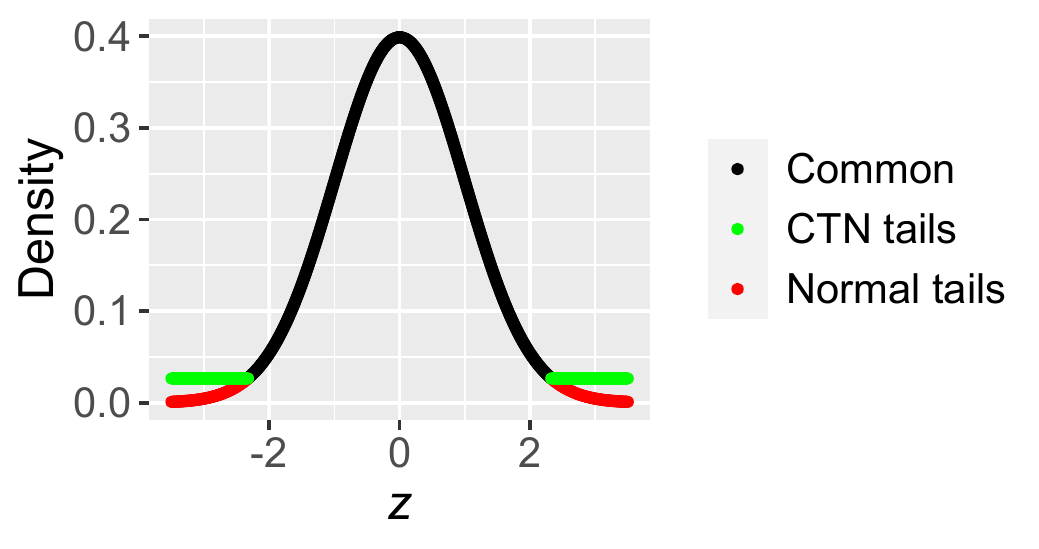} \cr
    \hspace{-20mm} \textbf{(a)} & \hspace{-20mm} \textbf{(b)}
\end{array}$
  \vspace{-3mm}
\caption{\small PDFs of the standard normal distribution and the CTN distribution with (a) $\varrho = 0.95$ and (b) $\varrho = 0.98$ }\label{fig:CTN}
\end{figure}

The distribution that is introduced to resolve a conflict due to either a prior location significantly different than that of the likelihood or small prior scalings is the CTN distribution. Its density is as follows:
 \begin{align}\label{def_CTN}
  g_{\text{CTN}}(z) := \left\{
                                                    \begin{array}{lcc}
                                                       \varphi(z)  & \text{ if } & |z| \leq \kappa, \\
                                                      \varphi(\kappa) & \text{ if } & |z|>\kappa,
                                                    \end{array}
\right.
  \end{align}
  where $z \in \re$ and $\kappa$ is a function of the sole free parameter of the CTN distribution, $\varrho \in (0, 1)$, with an analogous definition to $\tau$ in the previous section:
  \[
   \kappa = \Phi^{-1}((1+ \varrho)/2) = \{\kappa : \Prob(-\kappa \leq Z \leq \kappa) = \varrho \,\text{ for }\, Z\, \sim \, \mathcal{N}(0,1)\}.
  \]
  A CTN density with $\varrho = 0.95$ is presented in \autoref{fig:CTN} (a). We observe in \autoref{fig:CTN} (a) that even though the CTN density with $\varrho = 0.95$ matches the standard normal one on the same interval as the LPTN with $\rho = 0.95$ (\autoref{fig:LPTN} (a)), its level of similarity with the standard normal density is much lower. Increasing the value of $\varrho$ from $0.95$ to, for instance, $0.98$ alleviates this issue, as seen in \autoref{fig:CTN} (b), at the price of a slower conflict resolution (but not significantly slower as will be seen in \autoref{sec:simulation}). The effectiveness of CTN priors with $\varrho =0.98$ at resolving conflicts due to small prior scalings was shown in \autoref{fig:resolution} (b).

\normalsize

The main disadvantage of CTN prior distributions is that they are improper; they thus cannot be used when there are more parameters than observations. Another disadvantage is that the derivative of their densities is discontinuous, like that of LPTN densities. The discontinuity points of the CTN derivative have null measure, like those of the LPTN derivative, implying that samplers like HMC can be employed. The estimates and posterior variances needed for \autoref{sec:simulation} have been computed using HMC, and no problems have been encountered, like with the posterior distributions resulting from LPTN priors.

The main advantage of CTN prior distributions is their limiting behaviour: for any fixed $\beta_j$ and $\sigma$
\[
 \frac{\lambda_j}{\sigma} g_j\left(\frac{\lambda_j}{\sigma} (\beta_j - \mu_j) \right) \rightarrow \frac{\lambda_j}{\sigma} \varphi(\kappa) \propto \frac{1}{\sigma},
\]
whenever: i) $\mu_j \rightarrow \pm \infty$ and $\lambda_j$ is fixed, or ii) $\lambda_j \rightarrow \infty$ and $\mu_j$ is fixed but different than $\beta_j$. The conflict resolution is not perfect as the term $1 / \sigma$ does not disappear in the limiting posterior density. This term comes from the form of the prior (with a scale parameter given by $\sigma / \lambda_j$), and is not a consequence of insufficiently heavy tails as was the case for Student prior distributions for conflicts due to significantly different locations and LPTN prior distributions for conflicts due to small prior scalings. It should be seen as a flaw of the method. The tails of CTN densities are indeed sufficiently heavy, and allow to resolve any conflict that can happen under our regression framework. A consequence of their sufficiently heavy tails is that they yield a fast convergence towards the limiting posterior distribution, as will be seen in \autoref{sec:simulation}.

As mentioned for Student prior distributions which yield a similar trace (recall \eqref{eqn:limit_stu}), the remaining term $1 / \sigma$ for CTN prior distributions has an impact on the limiting posterior variability of all coefficients which is seen to be more or less significant depending on the sample size and the number of conflicting prior densities. However, contrarily to Student prior distributions, the impact does not increase with the level of similarity between CTN prior distributions and normal ones; recall that the trace left asymptotically by a conflicting Student prior distribution (due to significantly different locations) is $\sigma^\gamma$ and that the level of similarity with a normal prior is controlled through $\gamma$. The level of similarity between CTN prior distributions and normal ones is controlled through $\kappa$, and its value does not have an impact on the trace left asymptotically by a conflicting CTN prior distribution; the trace is $1 / \sigma$ regardless of the value of $\kappa$.

Note that, in an ideal setting where one knows how many conflicting prior distributions there are, one can multiply the prior of $\sigma$ (that would ideally be used in a situation where there is no conflict) by $\sigma$ with a power corresponding to the number of conflicting priors to perfectly resolve the conflict. In practice, one may have prior beliefs about that number, but cannot be sure about it. Consequently, we recommend to not alter the prior of $\sigma$ as it can cause more harm than good.

\section{Theoretical results}\label{sec:results}

In this section, we present three theoretical results. For the presentation of these results, it is required to introduce a proper mathematical framework and details about the model. This is done in \autoref{sec:ass_notation} and the results follow. In \autoref{sec:full_info}, we consider an ideal situation where one has access to full information about the conflict, namely, which prior distributions are in conflict and why. This situation is unrealistic but it allows to show what is an ideal conflict resolution in a regression framework. Next, in \autoref{sec:partial_info}, we present a result  in a situation where one has access to partial information, namely, that there is no conflict due to small prior scalings. This is a more realistic scenario that can be think of as one where a practitioner include information about the regression coefficients, but the practitioner is cautious while doing it, in the sense that the practitioner use moderate to large prior scalings. In practice (as we saw in \autoref{fig:conflicts} (a)), there is no certainty that there will be no conflict due to a prior location significantly different than that of the likelihood, even when using moderate to large prior scalings, and we consider that the practitioner wants to be protected against this risk. The last situation is the most common one where a practitioner wants to include information about the regression coefficients and thus set values for all $\mu_j$ and $\lambda_j$. While having no reason to believe \emph{a priori} that a conflict will occur (and thus while having no information regarding a potential conflict), the practitioner wants to be protected. A result in that situation is presented in \autoref{sec:no_info}.

Throughout the current section, we aim to characterize with theoretical results how conflicts are dealt with asymptotically when using heavy-tailed priors. Theoretical results like those in \cite{bunke1998asymptotic} allow to study the limiting behaviour of posterior distributions resulting from heavy-tailed priors under another asymptotic regime than that study here, namely the large-sample regime $n \rightarrow \infty$. Even if some heavy-tailed priors presented in \autoref{sec:heavy_priors} are non-smooth, it can be proved that they yield posterior distributions that concentrate around the correct parameter values as $n \rightarrow \infty$ and posterior estimates that are consistent and asymptotically normal, provided that the priors are non-conflicting ($\mu_j$ and $\lambda_j$ are all held fixed) and the data model is regular enough (which is the case, for instance, for linear regression and GLMs). This means that if there is no conflicting prior information, whether heavy-tailed priors are used or not  does not have an impact asymptotically as $n \rightarrow \infty$ on the posterior distributions and estimates.

\subsection{Mathematical framework}\label{sec:ass_notation}

We first precisely describe the asymptotic regime under which the theoretical results in the next subsections are stated. We assume that possibly some $\mu_j \rightarrow \pm \infty$ and/or some $\lambda_s \rightarrow \infty$ (with $s$ different than $j$). To analyse separately the effect of misspecified locations and scalings and to simplify the analysis, we indeed consider that when $\mu_j \rightarrow \pm \infty$, $\lambda_j$ is fixed, and when $\lambda_s \rightarrow \infty$, $\mu_s$ is fixed. We more precisely consider that for all $j$,
\begin{itemize}
    \itemsep 0mm

    \item $\mu_j = a_j + b_j \omega$, with $a_j, b_j \in \re$,

    \item $\lambda_j = c_j + d_j \omega$ with $c_j > 0$ and $d_j \geq 0$,
\end{itemize}
under the constraint that $b_j \neq 0$ for conflicting locations, but 0 otherwise, and $d_j > 0$ for conflicting scalings, but 0 otherwise, with $b_j = 0$ if $d_j > 0$ and $d_j = 0$ if $b_j \neq 0$, and we let $\omega \rightarrow \infty$. This framework allows, for instance, for conflicting scalings to decrease (because $\lambda_j \rightarrow \infty$) at different speeds, meaning that it represents situations where there may be several conflicting scalings, but their values, while being extreme, are not the same.

We now present the model assumptions and introduce required notation. Consider that we observed $n$ data points from a dependent variable, denoted by $y_1, \ldots, y_n \in \re$, where $n$ is a positive integer. Consider also that we have access to $n$ vectors of  $p \in \{2, 3, \ldots\}$ covariates, denoted by $\mathbf{x}_1 := (x_{11}, \ldots, x_{1p})^T, \ldots, \mathbf{x}_n := (x_{n1}, \ldots, x_{np})^T \in \re^p$, where in particular $x_{11} = \ldots = x_{n1} = 1$ to introduce an intercept in the model. As typically done in linear regression, we treat these vectors as known constants, i.e.\ not as realizations of random variables, contrarily to $y_1, \ldots, y_n$. The posterior distribution is thus conditional on the latter only.

In linear regression, the random variables $Y_i$ are modelled as $Y_i = \mathbf{x}_i^T \boldsymbol\beta + \sigma\varepsilon_i$, $i = 1,\ldots, n$, where $\varepsilon_1, \ldots, \varepsilon_n \in \re$ are random standardized errors. We assume that the $n + 2$ random variables  $\varepsilon_1, \ldots, \varepsilon_n$, $\boldsymbol\beta$ and $\sigma$ are independent, implying that
\[
 \varepsilon_i \mid \boldsymbol\beta, \sigma \stackrel{\mathcal{D}}{=} \varepsilon_i \sim f, \quad i = 1,\ldots, n,
\]
where ``$\, \stackrel{\mathcal{D}}{=} \,$'' denotes an equality in distribution. This latter assumption is common.

The resulting posterior density is given by
\begin{align}\label{eqn_post}
 \pi_\omega(\boldsymbol\beta, \sigma \mid \mathbf{y}) := \pi_\omega(\boldsymbol\beta, \sigma) \left[\prod_{i = 1}^n (1 / \sigma) f((y_i - \mathbf{x}_i^T \boldsymbol\beta) / \sigma)\right] \Bigg/ m_\omega(\mathbf{y}), \quad \boldsymbol\beta \in \re^p, \sigma > 0,
\end{align}
where $\mathbf{y} := (y_1, \ldots, y_n)^T$, $\pi_\omega(\, \cdot \,, \cdot \,)$ is the prior density and
\begin{align*}
 m_\omega(\mathbf{y}) :=  \int_{\re^p} \int_0^\infty \pi_\omega(\boldsymbol\beta, \sigma) \left[\prod_{i = 1}^n (1 / \sigma) f((y_i - \mathbf{x}_i^T \boldsymbol\beta) / \sigma)\right] \, \d\sigma \, \d\boldsymbol\beta.
\end{align*}
A dependence on $\omega$ (implying a potential presence of conflict) is highlighted using a subscript. The definition of the posterior distribution in \eqref{eqn_post} only makes sense when the density is integrable, and thus the marginal density $m_\omega(\mathbf{y})$, playing the role of a normalizing constant in this case, is finite. We provide in the next subsections sufficient conditions ensuring that this is the case for all $\omega$ and for the limiting posterior density. The limiting posterior distribution is denoted by $\overline{\pi}(\, \cdot \,, \cdot \mid \mathbf{y})$ and its normalizing contant is $\overline{m}$. Their expressions depend on the situations presented in the next subsections.

We now present regularity conditions on $f$. We assume that:
\begin{itemize}
 \itemsep 0mm
 \item $f$ is a strictly positive continuous PDF that is symmetric with respect to 0;
 \item all parameters of $f$, if any, are known;
 \item there exists a threshold above which the function $\xi$ defined by $z \mapsto z f(z)$ is monotonic;
 \item there exists a positive constant $M$ such that $f / g_{\text{LPTN}} \leq M$.
\end{itemize}
Examples of PDFs satisfying these conditions include those of normal, Laplace, Student (with pre-specified degrees of freedom) and LPTN (with pre-specified $\rho$) distributions. The last assumption above on $f$ is about the tail decay of $f$; it must be at most as slow as that of $g_{\text{LPTN}}$. This implies that our results are also valid when heavy-tailed error distributions are used for robustness against outliers.

The assumptions on $\pi_\omega(\, \cdot \mid \sigma)$ have been presented in \autoref{sec:context}. Denote by $\pi(\, \cdot \,)$ the prior of $\sigma$ that would ideally be used in a situation where there is no conflict. The assumptions on this density depend on the situations presented in the next subsections and will thus be stated in these subsections.

We finish this section by defining the index set of conflicting priors: $\text{C} := \{j: b_j \neq 0 \text{ or } d_j > 0\}$. The index set of non-conflicting priors is thus given by: $\text{C}^\mathsf{c}$. We also define two subsets of $\text{C}$: $\text{C}_{\text{b}} := \{j: b_j \neq 0\}$ and $\text{C}_{\text{d}} := \{j: d_j > 0\}$, which are such that $\text{C}_{\text{b}} \bigcup \text{C}_{\text{d}} = \text{C}$ and $\text{C}_{\text{b}} \bigcap \text{C}_{\text{d}} = \varnothing$.

\subsection{Full information}\label{sec:full_info}

Consider that we have set values for all $\mu_j$ and $\lambda_j$, and that we are provided with the set $\text{C}$. We use the latter to set all $g_j$ accordingly. More precisely, for all $j \in \text{C}_{\text{b}}$, we set $g_j = g_{\text{LPTN}}$, and for all $j \in \text{C}_{\text{d}}$, we set $g_j = g_{\text{CTN}}$. We consider that non-conflicting priors, with $j \in \text{C}^\mathsf{c}$, are set to proper distributions with densities having tails not more heavy than those of LPTN densities. Given that we are provided with the set $\text{C}$ and we set some priors to CTN distributions (if $\text{C}_{\text{d}} \neq \varnothing$), we adjust the prior on $\sigma$ to get rid of the trace left asymptotically by CTN distributions, i.e.\ the resulting prior density is proportional to $\pi(\sigma)$ multiplied by $\sigma^{|\text{C}_{\text{d}}|}$. We assume that $\pi(\sigma)$ is bounded above by a constant or a constant times $1/\sigma$, for all $\sigma > 0$, which allows for most proper prior distributions and improper prior densities proportional to $1/\sigma$ or $1$.

\begin{Theorem}\label{thm_conv_1}
  Assume that for all $j \in \text{C}_{\text{b}}$, $g_j = g_{\text{LPTN}}$, for all $j \in \text{C}_{\text{d}}$, $g_j = g_{\text{CTN}}$, and for all $j \in \text{C}^\mathsf{c}$, the positive constant $M$ can be chosen such that $g_j / g_{\text{LPTN}} \leq M$. Assume that the prior on $\sigma$ has a density that is proportional to $\sigma^{|\text{C}_{\text{d}}|} \pi(\sigma)$, for all $\sigma > 0$. Assume that the constant $M$ can be chosen such that $\pi(\sigma) \leq \max(M, \sigma^{-1} M)$. Assume that $n + |\text{C}^\mathsf{c}| \geq 2p - 1 + |\text{C}_{\text{b}}|$. Under the framework described in \autoref{sec:ass_notation} (recall in particular the form of the prior distribution \eqref{eqn:prior}, the definition of the posterior distribution \eqref{eqn_post}, and that $\mu_j = a_j + b_j \omega$ and $\lambda_j = c_j + d_j \omega$), and as $\omega \rightarrow \infty$,
%
%
%
%
 the posterior distribution converges:
 \[
  \pi_\omega(\, \cdot \,, \cdot \mid \mathbf{y}) \rightarrow \overline{\pi}(\, \cdot \, , \cdot \mid \mathbf{y}),
 \]
 where
 \begin{align}\label{eqn:post_1}
  \overline{\pi}(\boldsymbol\beta, \sigma \mid \mathbf{y}) := \pi(\sigma) \prod_{j \in \text{C}^\mathsf{c}} \pi_j(\beta_j \mid \sigma) \left[\prod_{i = 1}^n (1 / \sigma) f((y_i - \mathbf{x}_i^T \boldsymbol\beta) / \sigma)\right] \Bigg/ \overline{m}(\mathbf{y}), \cr
  \boldsymbol\beta \in \re^p, \sigma > 0,
 \end{align}
 with
 \begin{align*}
  \overline{m}(\mathbf{y}) = \int_{\re^p} \int_0^\infty \pi(\sigma) \prod_{j \in \text{C}^\mathsf{c}} \pi_j(\beta_j \mid \sigma) \left[\prod_{i = 1}^n (1 / \sigma) f((y_i - \mathbf{x}_i^T \boldsymbol\beta) / \sigma)\right] \, \d\sigma \, \d\boldsymbol\beta.
 \end{align*}

\end{Theorem}

The result of \autoref{thm_conv_1} essentially follows from a characterization of the asymptotic behaviour of the marginal distribution:
 \begin{align}\label{eq:conv_marg}
  \frac{m_\omega(\mathbf{y})}{\prod_{j \in \text{C}_{\text{b}}} g_j(\mu_j) \prod_{j \in \text{C}_{\text{d}}} \lambda_j \, g_j(\kappa)} \rightarrow \overline{m}(\mathbf{y}),
 \end{align}
 with $m_\omega(\mathbf{y}) / [\prod_{j \in \text{C}_{\text{b}}} g_j(\mu_j) \prod_{j \in \text{C}_{\text{d}}} \lambda_j \, g_j(\kappa)] < \infty$ and $\overline{m}(\mathbf{y})< \infty$ (implying that the posterior distributions are proper); recall that $\kappa = \Phi^{-1}((1+ \varrho)/2)$, where $\varrho$ is the parameter of the CTN distribution. From the characterization in \eqref{eq:conv_marg} we can, indeed, prove that the posterior density converges pointwise, which in turn allows to prove the convergence of the posterior distribution using Scheffé's theorem (see \cite{scheffe1947useful}).

 To prove \eqref{eq:conv_marg}, we exploit the proof of Theorem 2.1 in \cite{gagnon2020}. That paper is about robustness to outliers in linear regression. Theorem 2.1 in \cite{gagnon2020} characterizes the limiting behaviour of the posterior distribution as some $y_i \rightarrow \pm \infty$. The prior on all parameters is assumed to be non-conflicting with a joint prior density bounded above by $\max(M, \sigma^{-1} M)$. To exploit the proof of that result in \cite{gagnon2020}, we write
\[
 \frac{m_\omega(\mathbf{y})}{\prod_{j \in \text{C}_{\text{b}}} g_j(\mu_j) \prod_{j \in \text{C}_{\text{d}}} \lambda_j \, g_j(\kappa) \, \overline{m}(\mathbf{y})}
\]
as an integral that is seen to converge to 1 if we are allowed to interchange the limit $\omega \rightarrow \infty$ and the integral. We verify that we are allowed to do this by using Lebesgue's dominated convergence theorem. The problem then becomes to prove that the integrand is bounded by an integrable function of $\boldsymbol\beta$ and $\sigma$ that does not depend on $\omega$. This is the main difficulty. We show that it is sufficient to bound above
\begin{align}\label{eqn:fct_thm_1}
  \overline{\pi}(\boldsymbol\beta, \sigma \mid \mathbf{y}) \left[\prod_{j \in \text{C}_{\text{b}}} \frac{\pi_{j, \omega}(\beta_j \mid \sigma)}{g_j(\mu_j)}\right] = \overline{\pi}(\boldsymbol\beta, \sigma \mid \mathbf{y}) \left[\prod_{j \in \text{C}_{\text{b}}} \frac{\frac{\lambda_j}{\sigma} g_j\left(\frac{\lambda_j}{\sigma} (\beta_j - \mu_j) \right)}{g_j(\mu_j)}\right]
\end{align}
 by an integrable function of $\boldsymbol\beta$ and $\sigma$ that does not depend on $\omega$.

 To fit within the framework of \cite{gagnon2020}, it suffices to treat $\lambda_j \mu_j$ for $j \in \text{C}^\mathsf{c} \bigcup \text{C}_{\text{b}}$ as an observation from the dependent variable and $\lambda_j$ that multiplies $\beta_j$ for $j$ in the same set as a vector of covariates where the other covariates are all equal to 0. Then we realize that a technical and lengthy part of the proof of Theorem 2.1 in \cite{gagnon2020} is devoted to a proof that a function of which \eqref{eqn:fct_thm_1} is a special case is bounded by an integrable function of $\boldsymbol\beta$ and $\sigma$ that does not depend on $\omega$. The main challenge is that the terms $g_j(\mu_j)$ in the denominator of the product in \eqref{eqn:fct_thm_1} goes to 0 as $\omega \rightarrow \infty$; the strategy is thus to find a way to get rid of these terms by finding an upper bound for any $(\boldsymbol\beta, \sigma)$.

 When $\beta_j$ is far from $\mu_j$, we can use \autoref{prop:location-scale} to bound $\pi_{j, \omega}(\beta_j \mid \sigma)/g_j(\mu_j)$ in \eqref{eqn:fct_thm_1}. But this does not work when $\beta_j$ is not far from $\mu_j$. In this case, we have to use a density $(1 / \sigma) f((y_i - \mathbf{x}_i^T \boldsymbol\beta) / \sigma)$ in $\overline{\pi}(\boldsymbol\beta, \sigma \mid \mathbf{y})$ which is presumably close to 0 when $\beta_j$ is not far from $\mu_j \rightarrow \pm \infty$, and bound above $(1 / \sigma) f((y_i - \mathbf{x}_i^T \boldsymbol\beta) / \sigma) / g_j(\mu_j)$. Note that we can also use a prior density with $j \in \text{C}^\mathsf{c}$. The proof is based on a decomposition of the parameter space into disjoint sets; for each of these sets, we are able to identify in which case we precisely are. In the case where $\beta_j$ is not far from $\mu_j$, it is shown that the associated hyperplanes pass close to at most $p - 1$ non-conflicting sources (data points in the case of  \cite{gagnon2020}) using that $\mathbf{x}_i$ can be written as a linear combination of $p$ other covariate vectors and the explicit form of the linear-regression model. Other non-conflicting data points are thus such that $(1 / \sigma) f((y_i - \mathbf{x}_i^T \boldsymbol\beta) / \sigma)$ are close to 0. The argument is technical and essentially consists in isolating cases where the parameters are such that the densities of \emph{conflicting} sources are evaluated in the tails and those where the parameters are instead such that the densities of \emph{non-conflicting} sources are evaluated in the tails; there is no reason to believe that the result does not hold for other regression models, and in particular, in the general regression framework presented in \autoref{sec:context} including GLMs, perhaps under different assumptions. We believe that even though it turns out that the assumptions are indeed different, they will be similar in essence.

 The assumption that $n + |\text{C}^\mathsf{c}| \geq 2p - 1 + |\text{C}_{\text{b}}|$ is essentially to ensure that the non-conflicting sources of information are dominant. It is a consequence of: when $\beta_j$ is not far from $\mu_j$, possibly $|\text{C}_{\text{b}}|$ non-conflicting sources are required to get rid of terms $g_j(\mu_j)^{-1}$ in \eqref{eqn:fct_thm_1}, and imagine that the number of non-conflicting sources left is $2p - 1$, then $p - 1$ of them may be close to hyperplanes such that $\beta_j$ is not far from $\mu_j$ and, using the decomposition in \cite{gagnon2020}, it is shown that $p$ non-conflicting sources are sufficient to obtain an integrable function.

 By looking at \eqref{eqn:post_1}, we see that we get rid asymptotically of all the conflicting priors and no trace is left; the resulting limiting posterior distribution is that with improper Jeffreys priors $\pi_j(\beta_j \mid \sigma) \propto 1$, for $j \in \text{C}$. We thus have a characterization of the limiting behaviour of the posterior distribution/density and estimates like maximum a posteriori probability (MAP) estimates and posterior medians. It is possible to show under additional mild assumptions that the posterior expectations and the joint posterior distribution of a model indicator and parameters in a context of variable selection converge as well. All these results thus characterize the limiting behaviour of a variety of Bayes estimators. Analogous results hold in the situations that are presented in Sections \ref{sec:partial_info} and \ref{sec:no_info}.

\subsection{Partial information}\label{sec:partial_info}

Now, consider that we have set values for all $\mu_j$ and $\lambda_j$, and we know that $\text{C}_{\text{d}} = \varnothing$. In practice, the situation is rather that a practitioner is confident that there will be no conflict due to small scalings. We now describe how to set the priors in this case and the limiting behaviour of the posterior distribution if the practitioner turns out to be right. Given that each of the priors on the regression coefficients is exposed to a potential conflict due to a prior location significantly different than that of the likelihood, we set $g_j = g_{\text{LPTN}}$ for all $j$. The advantage here is that, because no CTN distribution is used, no adjustment on the prior of $\sigma$ is required to yield, as in the previous section, a limiting posterior distribution without a trace of conflict and with improper Jeffreys priors $\pi_j(\beta_j \mid \sigma) \propto 1$, for $j \in \text{C}$. The prior on $\sigma$ is thus set to $\pi(\, \cdot \,)$ and we assume that $\pi(\sigma)$ is bounded above by a constant or a constant times $1/\sigma$, for all $\sigma > 0$, as before.

\begin{Theorem}\label{thm:conv_2}
  Assume that $\text{C}_{\text{d}} = \varnothing$. Assume that for all $j$, $g_j = g_{\text{LPTN}}$. Assume that the prior on $\sigma$ is $\pi(\, \cdot \,)$ and that $\pi(\sigma) \leq \max(M, \sigma^{-1} M)$ for all $\sigma > 0$. Assume that $n + |\text{C}^\mathsf{c}| \geq 2p - 1 + |\text{C}_{\text{b}}|$. Under the framework described in \autoref{sec:ass_notation} (recall in particular the form of the prior distribution \eqref{eqn:prior}, the definition of the posterior distribution \eqref{eqn_post}, and that $\mu_j = a_j + b_j \omega$ and $\lambda_j = c_j + d_j \omega$), and as $\omega \rightarrow \infty$,
%
%
%
 the posterior distribution converges:
 \[
  \pi_\omega(\, \cdot \,, \cdot \mid \mathbf{y}) \rightarrow \overline{\pi}(\, \cdot \, , \cdot \mid \mathbf{y}),
 \]
 where $\overline{\pi}(\, \cdot \, , \cdot \mid \mathbf{y})$ is defined as in \eqref{eqn:post_1}.
%

\end{Theorem}

\autoref{thm:conv_2} is an adaptation of \autoref{thm_conv_1} in which it is considered that $\text{C}_{\text{d}} = \varnothing$, which implies that $\text{C}_{\text{b}} = \text{C}$. Also, the proof of \autoref{thm:conv_2} is an adaptation of that of \autoref{thm_conv_1}. For the same reasons as those explained in \autoref{sec:full_info}, we thus believe that \autoref{thm:conv_2} holds in the general regression framework presented in \autoref{sec:context} including GLMs, perhaps under different, yet similar, assumptions. A difference between \autoref{thm:conv_2} and \autoref{thm_conv_1} is that, because we do not know which of the priors will be in conflict (if any) and thus set all $g_j = g_{\text{LPTN}}$, the prior distributions in the limiting posterior for $j \in \text{C}^\mathsf{c}$ are thus all LPTN distributions; in \autoref{thm_conv_1}, they can be selected to be otherwise, provided that they are proper distributions with densities having tails not more heavy than those of LPTN densities. Using LPTN prior distributions is perhaps not the first choice for a practitioner, but this comes with protection, as seen in \autoref{thm:conv_2}.

It is possible to prove a similar result to \autoref{thm:conv_2} if instead we set $g_j$ to a Student distribution for each $j$. The difference is that the limiting posterior distribution is defined otherwise than in \eqref{eqn:post_1}. It is instead such that
\[
 \overline{\pi}(\boldsymbol\beta, \sigma \mid \mathbf{y}) \propto \sigma^{|\text{C}_{\text{b}}| \gamma} \pi(\sigma) \prod_{j \in \text{C}^\mathsf{c}} \pi_j(\beta_j \mid \sigma) \left[\prod_{i = 1}^n (1 / \sigma) f((y_i - \mathbf{x}_i^T \boldsymbol\beta) / \sigma)\right],
\]
reflecting that Student distributions asymptotically leave a trace in case of conflict, namely $\sigma^{\gamma}$ for each of the conflicting priors, and thus that they only partially resolves conflicts due to significantly different locations.

\subsection{No information}\label{sec:no_info}

In the last scenario, we consider that after setting all $\mu_j$ and $\lambda_j$, we have no reason to believe that these choices of locations and scalings will create conflicts, but we want to be protected in case it happens. We thus set $g_j = g_{\text{CTN}}$ for all $j$ to be prepared for all eventualities. As mentioned, the main disadvantage of using CTN priors is that they are improper. Their densities do not integrate and each $g_j$ is multiplied by $\lambda_j / \sigma$ to yield $\pi_{j, \omega}(\beta_j \mid \sigma)$ (recall \eqref{eqn:prior}). This implies that these densities cannot be used to integrate over $\boldsymbol\beta$ when verifying, for instance, that $\pi_\omega(\, \cdot \,, \cdot \mid \mathbf{y})$ is proper; the best that can be done is to bound them by a constant (that possibly depends on $\omega$) that is multiplied by $\sigma^{-p}$, and to use the (conditional) densities of $Y_1, \ldots, Y_n$ to integrate over  $\boldsymbol\beta$ (requiring $n \geq p$). Therefore, in order to obtain a proper posterior distribution, the prior on $\sigma$ needs to be such that $\int \sigma^{-p} \, \pi(\sigma) \, \d\sigma < \infty$. The good news is that setting $\pi(\, \cdot \,)$ such that $\sigma^2$ has an inverse-gamma distribution, as often done in practice \citep{1984west431, raftery1997bayesian}, implies that $\int \sigma^{-p} \, \pi(\sigma) \, \d\sigma < \infty$ for any $p$, and choice of shape and scale parameters for the inverse-gamma distribution.

\begin{Theorem}\label{thm:conv_3}
  Assume that for all $j$, $g_j = g_{\text{CTN}}$. Assume that the prior on $\sigma$ is $\pi(\, \cdot \,)$ and that it is such that $\int \sigma^{-p} \, \pi(\sigma) \, \d\sigma < \infty$. Assume that $n \geq p$. Under the framework described in \autoref{sec:ass_notation} (recall in particular the form of the prior distribution \eqref{eqn:prior}, the definition of the posterior distribution \eqref{eqn_post}, and that $\mu_j = a_j + b_j \omega$ and $\lambda_j = c_j + d_j \omega$), and as $\omega \rightarrow \infty$,
%
%
%
 the posterior distribution converges:
 \[
  \pi_\omega(\, \cdot \,, \cdot \mid \mathbf{y}) \rightarrow \overline{\pi}(\, \cdot \, , \cdot \mid \mathbf{y}),
 \]
 where
  \begin{align}\label{eqn:post_3}
  \overline{\pi}(\boldsymbol\beta, \sigma \mid \mathbf{y}) := \sigma^{-|\text{C}|} \, \pi(\sigma) \prod_{j \in \text{C}^\mathsf{c}} \pi_j(\beta_j \mid \sigma) \left[\prod_{i = 1}^n (1 / \sigma) f((y_i - \mathbf{x}_i^T \boldsymbol\beta) / \sigma)\right] \Bigg/ \overline{m}(\mathbf{y}),
 \end{align}
 $\boldsymbol\beta \in \re^p, \sigma > 0$, with
 \begin{align*}
  \overline{m}(\mathbf{y}) = \int_{\re^p} \int_0^\infty \sigma^{-|\text{C}|} \, \pi(\sigma) \prod_{j \in \text{C}^\mathsf{c}} \pi_j(\beta_j \mid \sigma) \left[\prod_{i = 1}^n (1 / \sigma) f((y_i - \mathbf{x}_i^T \boldsymbol\beta) / \sigma)\right] \, \d\sigma \, \d\boldsymbol\beta.
 \end{align*}
%

\end{Theorem}

The proof of \autoref{thm:conv_3} is much simpler than those of Theorems \ref{thm_conv_1} and \ref{thm:conv_2}. While still using Lebesgue's dominated convergence theorem, the term that is sufficient to bound by an integrable function of $\boldsymbol\beta$ and $\sigma$ that does not depend on $\omega$ is
\[
 \overline{\pi}(\boldsymbol\beta, \sigma \mid \mathbf{y}) \left[\prod_{j \in \text{C}} \frac{g_{\text{CTN}}\left(\frac{\lambda_j}{\sigma} (\beta_j - \mu_j) \right)}{g_{\text{CTN}}(\kappa)} \right]  \leq  \overline{\pi}(\boldsymbol\beta, \sigma \mid \mathbf{y}) \, M^{|\text{C}|},
\]
which is thus bounded by a function that does not depend on $\omega$, contrarily to \eqref{eqn:fct_thm_1}. The core of the proof of \autoref{thm:conv_3} is essentially devoted to proving that $\overline{\pi}(\, \cdot \,, \, \cdot \mid \mathbf{y})$ is proper (requiring $n \geq p$ and that $\int \sigma^{-p} \, \pi(\sigma) \, \d\sigma < \infty$ in our case, as seen in \autoref{thm:conv_3}). In other words, \autoref{thm:conv_3} holds for any of the regression model fitting in the general regression framework presented in \autoref{sec:context} including GLMs, provided that the prior distribution exhibit a conditional-independence structure as in \eqref{eqn:prior} and the resulting limiting posterior distribution \eqref{eqn:post_3} is proper. With \autoref{thm:conv_3}, it is thus even clearer than with Theorems \ref{thm_conv_1} and \ref{thm:conv_2} that the result holds in the general regression framework presented in \autoref{sec:context}, perhaps under different, yet similar, assumptions.

As mentioned previously, a weakness of using CTN prior distributions is that a trace asymptotically remains in case of conflict, namely $\sigma^{-|\text{C}|}$, as seen in \eqref{eqn:post_3}. This is similar to what happens when using Student prior distributions (recall the discussion at the end of \autoref{sec:partial_info}). However, there is an important difference: CTN prior distributions are effective against all types of conflicting situations, including those due to conflicting prior scalings, contrarily to Student prior distributions. Also, the degree of discrepancy between the resulting limiting posterior distribution and the ideal one (obtained in Theorems \ref{thm_conv_1} and \ref{thm:conv_2}), measured through the exponent of $\sigma$, does not depend on the level of similarity between the CTN distributions used and the standard normal (measure through the parameter $\varrho$). With Student prior distributions, the degree of discrepancy between the resulting limiting posterior distribution and the ideal one depends on the degrees of freedom $\gamma$. Recall that we recommend to not alter the prior of $\sigma$ with the aim of correcting for a discrepancy because we do not know $|\text{C}|$ \emph{a priori} and thus an adjustment can cause more harm than good.

\section{Simulation study}\label{sec:simulation}

A goal with this section is to show the impact of using an informative prior instead of a non-informative one, especially in the situation where the former is conflicting. Another goal is to identify suitable values for the hyperparameters of the heavy-tailed priors. We achieve all that through a simulation study; it suggests that $\gamma = 4$ degrees of freedom for Student prior distributions, $\rho = 0.95$ for LPTN prior distributions and $\varrho = 0.98$ for CTN prior distributions are suitable values. For the simulation study, we consider the normal-linear-regression framework, i.e.\ $Y_i = \mathbf{x}_i^T \boldsymbol\beta + \sigma \varepsilon_i$ with $f = \mathcal{N}(0, 1)$. For the reasons mentioned in \autoref{sec:results}, we expect the results to be similar in other regression frameworks, such as with GLMs. To simplify, we consider that the covariates are orthogonal and that the variables are standardized, i.e.\ $(1/n) \sum_{i=1}^n y_{i} = 0$ and $(1/n) \sum_{i=1}^n y_{i}^2 = 1$, $(1/n) \sum_{i=1}^n x_{ij} = 0$ and $(1/n) \sum_{i=1}^n x_{ij}^2 = 1$ for all $j$ (except for $j = 1$ for which $(1/n) \sum_{i=1}^n x_{i1} = 1$), and $\sum_{i=1}^n x_{ij} x_{is} = 0$ for $j \neq s$. Under this framework, the likelihood function exhibits a hierarchical and product form and is proportional to:
\[
 \frac{1}{\sigma^{n - p}}\exp\left(-\frac{1}{2 \sigma^2} \|\mathbf{y} - \hat{\mathbf{y}}\|_2^2\right)\prod_{j = 1}^p \frac{1}{\sigma}\exp\left(-\frac{n}{2\sigma^2}(\beta_j - \hat{\beta}_j)^2\right),
\]
where $\| \, \cdot \, \|_2$ is the Euclidian norm and $\hat{\mathbf{y}} := \mathbf{X} \hat{\boldsymbol\beta}$, $\mathbf{X}$ being the design matrix and $\hat{\boldsymbol\beta} := (\hat{\beta}_1, \ldots, \hat{\beta}_p)^T$ the OLS estimate, which in this case is such that $\hat{\beta}_j = (1 / n) \sum_{i=1}^n x_{ij} y_i$.

With this likelihood form, setting any prior $\pi_{j, \omega}(\, \cdot \mid \sigma)$ on $\beta_j$ with $\mu_j = \hat{\beta}_j$ yields the same marginal posterior distributions of the other coefficients regardless of the values of $\hat{\beta}_j$ and $\lambda_j$, as long as $\hat{\mathbf{y}}$ is the same. To simplify, we consider that $\mu_j = \hat{\beta}_j$ for all coefficients except one, namely $\beta_2$, that will be used to show the impact of different choices for $\mu_2, \lambda_2$ and $g_2$ to achieve our aforementioned goals. We also consider to simplify that $\hat{\boldsymbol\beta} = \mathbf{0}$, so that the marginal posterior distribution of $\beta_2$ only depends on $n$ (not on $p$ and the covariate data points); we set $n = 100$. It can be readily verified that $n > 3$ is sufficient to ensure a proper posterior distribution, even if the prior distributions of $\beta_2$ and $\sigma$ are improper Jeffreys priors. This condition is satisfied in the simulation study, and thus to simplify, we set the prior on $\sigma$ to the Jeffreys prior: $\pi(\sigma) \propto 1 / \sigma$. The non-informative Jeffreys prior on $\beta_2$ will serve as a benchmark, i.e.\ $\pi_{2, \omega}(\beta_2 \mid \sigma) \propto 1$, implying a posterior mean and variance of $0$ and $1 / (n - 3)$, respectively.

\begin{figure}[ht]
\centering
$\begin{array}{cc}
    \vspace{-2mm}\hspace{-2mm} \includegraphics[width=0.51\textwidth]{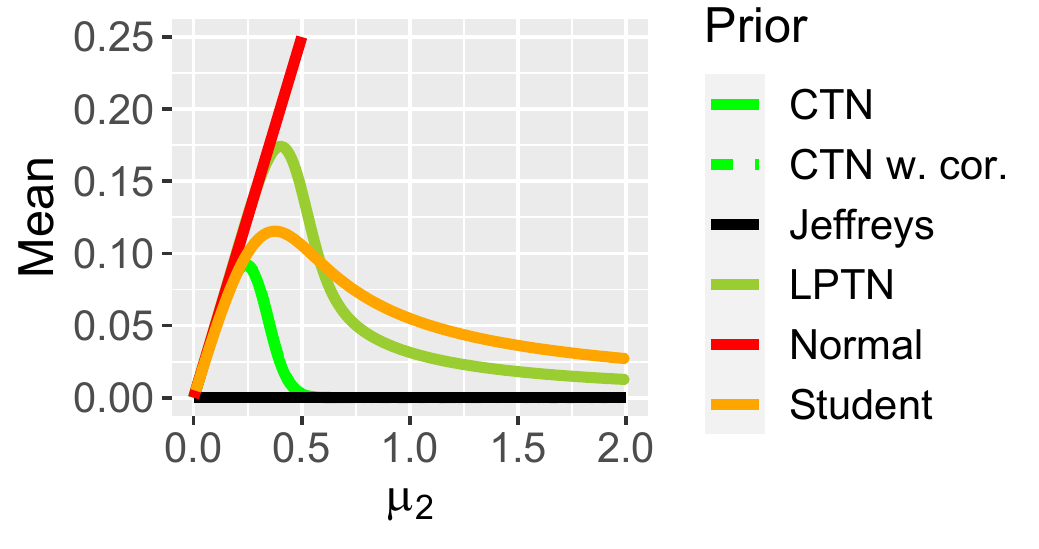} & \hspace{-5mm} \includegraphics[width=0.51\textwidth]{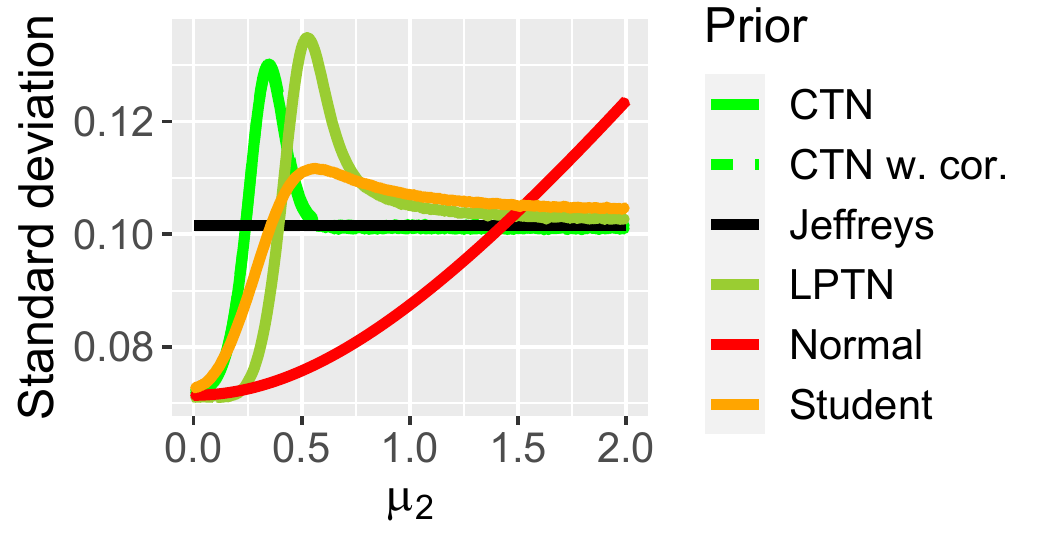} \cr
    \vspace{5mm}\hspace{-13mm} \textbf{(a) Mean as a function of $\mu_2$} & \hspace{-16mm} \textbf{(b) SD as a function of $\mu_2$} \cr
    \vspace{-2mm}\hspace{-2mm} \includegraphics[width=0.51\textwidth]{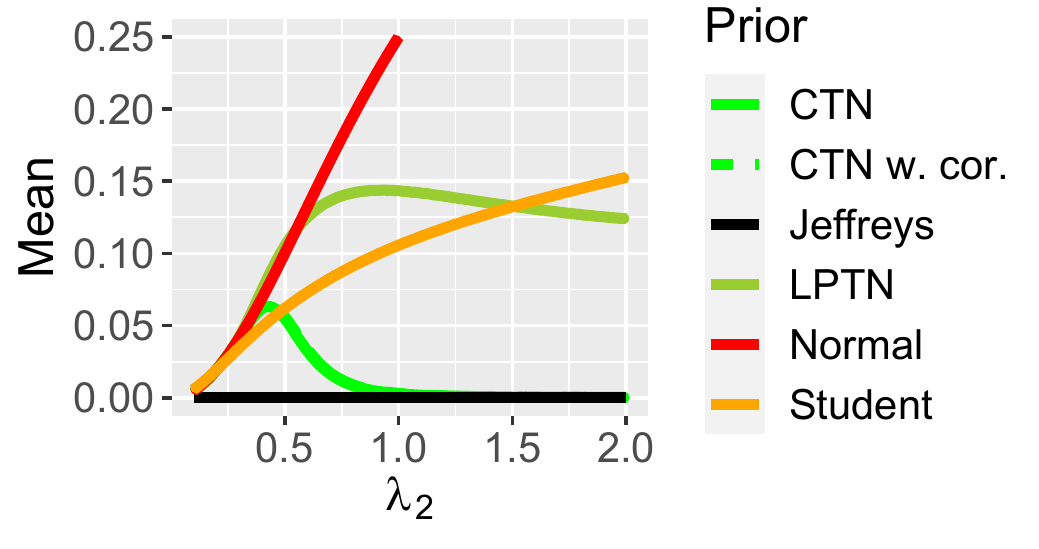} & \hspace{-5mm} \includegraphics[width=0.51\textwidth]{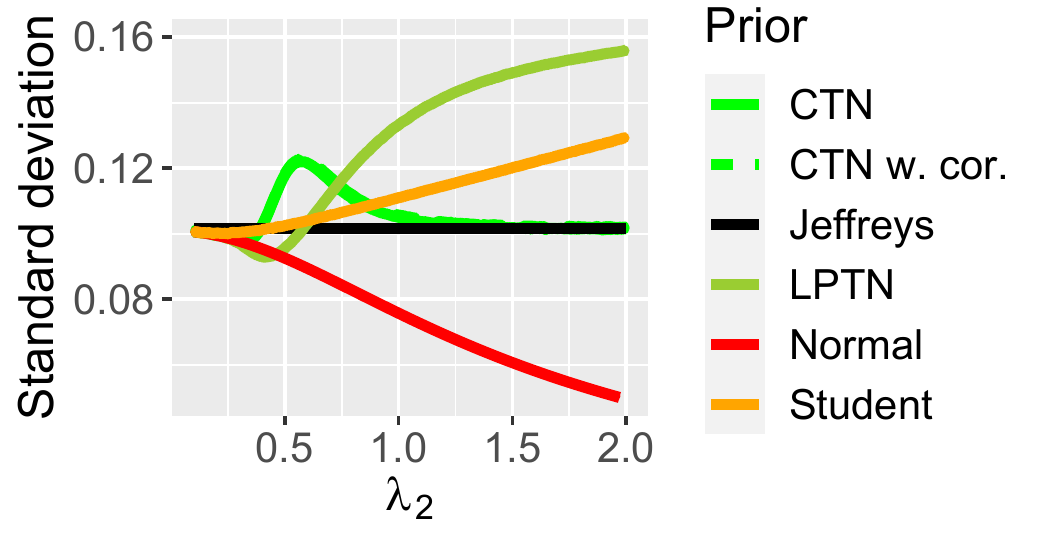} \cr
    \hspace{-13mm} \textbf{(c) Mean as a function of $\lambda_2$} & \hspace{-16mm} \textbf{(d) SD as a function of $\lambda_2$}
\end{array}$
  \vspace{-2mm}
\caption{\small Impact on posterior means and standard deviations as $\mu_2$ and $\lambda_2$ vary when a Jeffreys prior is used, i.e.\ $\pi_{2, \omega}(\beta_2 \mid \sigma) \propto 1$ (black line), and when $g_2$ is a standard normal (red line), a Student with $\gamma = 4$ degrees of freedom (orange line), a LPTN with $\rho = 0.95$ (dark green line), a CTN with $\varrho = 0.98$ and a CTN with $\varrho = 0.98$ but where the prior density of $\sigma$ is proportional to $\sigma \pi(\sigma)$ (both lines are light green, one is dashed, while the other one not; they are on top of each other); here SD stands for standard deviation}\label{fig:vary_mu and_lambda}
\end{figure}

\begin{figure}[ht]
\centering
$\begin{array}{ccc}
    \vspace{-2mm}\hspace{-3mm} \includegraphics[width=0.345\textwidth]{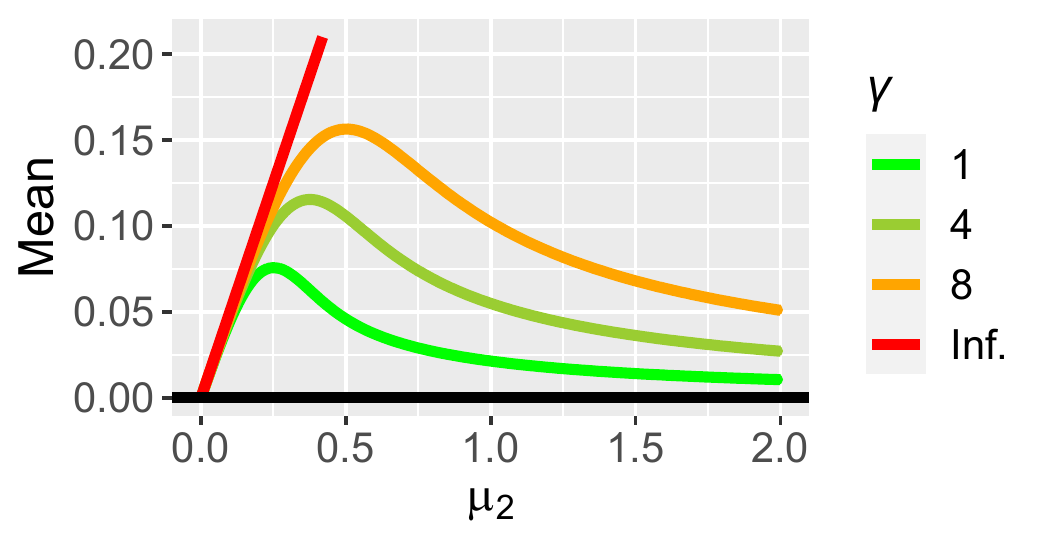} & \hspace{-5mm} \includegraphics[width=0.345\textwidth]{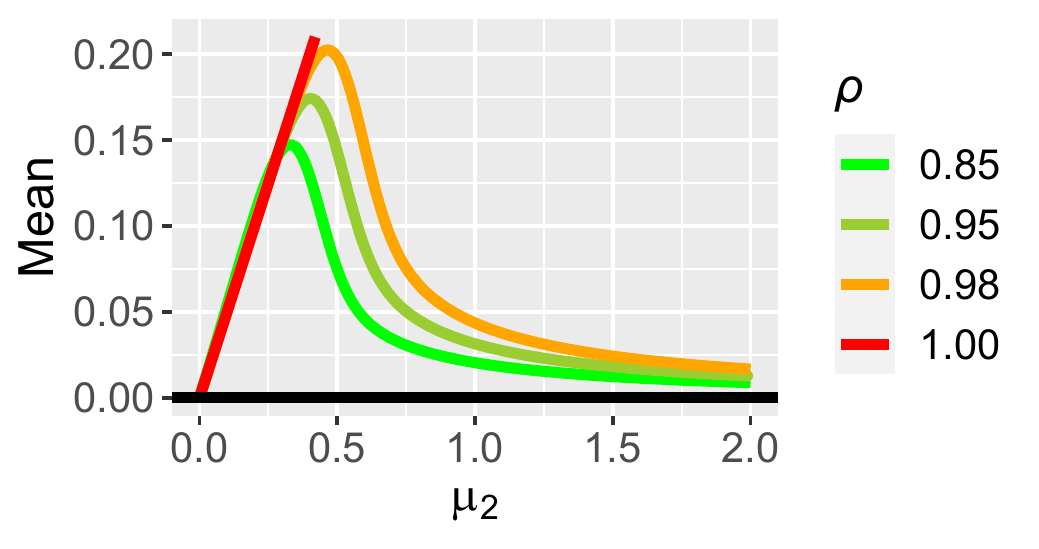} & \hspace{-5mm} \includegraphics[width=0.345\textwidth]{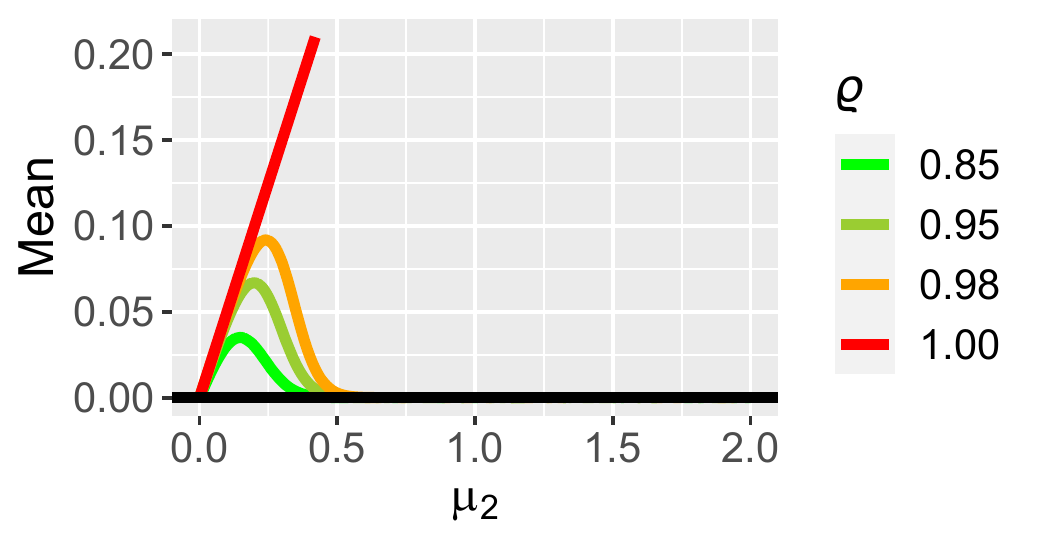}  \cr
    \vspace{-1mm}\hspace{-5mm} \textbf{(a) Mean as a func. of $\mu_2$} & \hspace{-5mm} \textbf{(b) Mean as a func. of $\mu_2$} & \hspace{-5mm} \textbf{(c) Mean as a func. of $\mu_2$} \cr
    \vspace{5mm}\hspace{-5mm} \textbf{when $g_2$ is a Student} & \hspace{-5mm} \textbf{when $g_2$ is a LPTN} & \hspace{-5mm} \textbf{when $g_2$ is a CTN} \cr
        \vspace{-2mm}\hspace{-3mm} \includegraphics[width=0.345\textwidth]{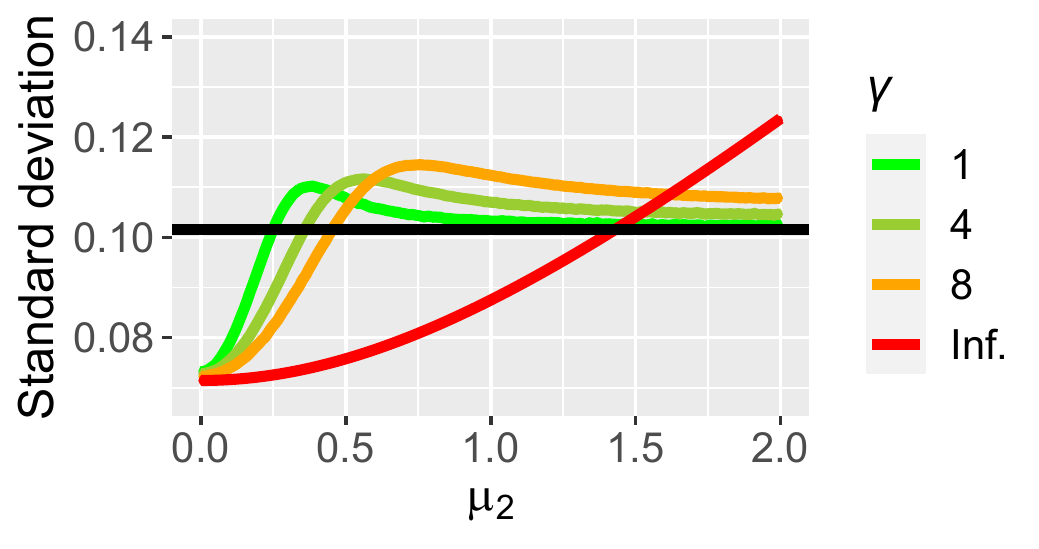} & \hspace{-5mm} \includegraphics[width=0.345\textwidth]{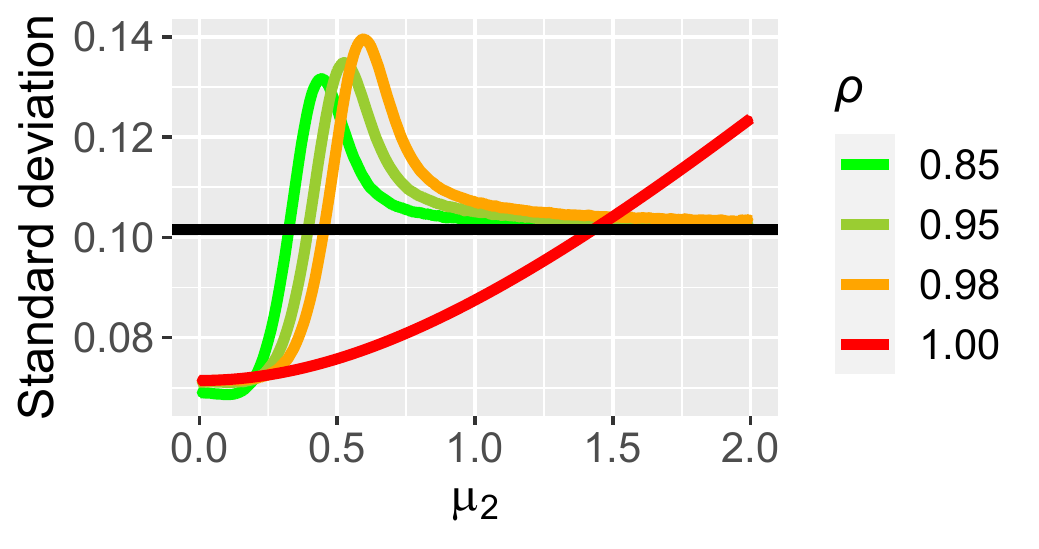} & \hspace{-5mm} \includegraphics[width=0.345\textwidth]{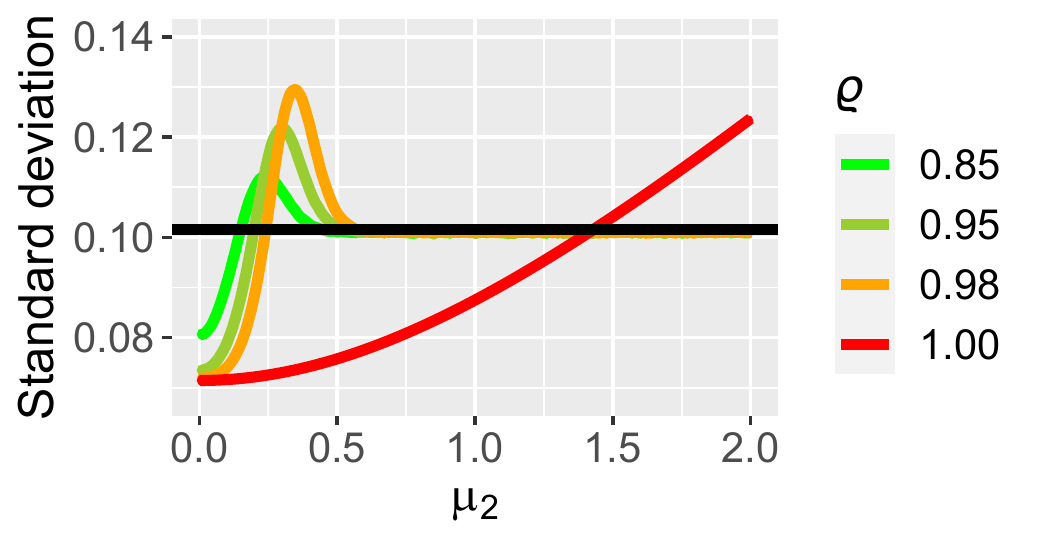}  \cr
   \vspace{-1mm}\hspace{-5mm} \textbf{(d) SD as a func. of $\mu_2$} & \hspace{-5mm} \textbf{(e) SD as a func. of $\mu_2$} & \hspace{-5mm} \textbf{(f) SD as a func. of $\mu_2$} \cr
   \hspace{-5mm} \textbf{when $g_2$ is a Student} & \hspace{-5mm} \textbf{when $g_2$ is a LPTN} & \hspace{-5mm} \textbf{when $g_2$ is a CTN} \cr
\end{array}$
  \vspace{-2mm}
\caption{\small Impact on posterior means and standard deviations as $\mu_2$ varies when: (a) and (d) $g_2$ is a Student, for different values of $\gamma$ (Inf. represents the standard normal); (b) and (e) $g_2$ is a LPTN, for different values of $\rho$ ($1.00$ represents the standard normal); (c) and (f) $g_2$ is a CTN, for different values of $\varrho$ ($1.00$ represents the standard normal); here SD stands for standard deviation and the black lines are again the results for the Jeffreys prior}\label{fig:vary_mu_mean_sd_hyper}
\end{figure}

We now describe the simulation study.
\begin{itemize}
 \itemsep 0mm

 \item The prior density on $\beta_2$ (except for the benchmark) is such that
 \[
  \pi_{2, \omega}(\beta_2 \mid \sigma) = \frac{\lambda_2 n^{1/2}}{\sigma}g_2\left(\frac{\lambda_2 n^{1/2}}{\sigma}(\beta_2 - \mu_2)\right).
 \]
 We present the results for 4 choices of informative $g_2$: a standard normal distribution, a Student distribution, a LPTN distribution and a CTN distribution. We compare them with one another and to the non-informative prior.

 \item While keeping $\lambda_2$ fixed and equal to 1, we gradually increase $\mu_2$ from 0 to 2. With this choice of $\lambda_2$, when $\mu_2 = 0$ the prior carries essentially the same information as the likelihood. We show the impact of more diffuse priors next. The results are presented in Figures \ref{fig:vary_mu and_lambda} (a)-(b) and \ref{fig:vary_mu_mean_sd_hyper}. Note that we observe similar results when considering a larger prior scaling, but we need to use an interval for $\mu_2$ with a larger upper bound.

 \item While keeping $\mu_2$ fixed and equal to 0.5 (to be able to appreciate a difference in location when the prior scaling conflicts), we gradually increase $\lambda_2$ from (nearly) 0 to 2. The results are presented in Figures \ref{fig:vary_mu and_lambda} (c)-(d).

\end{itemize}

\autoref{fig:vary_mu and_lambda} is used to compare the results produced by using different priors, while \autoref{fig:vary_mu_mean_sd_hyper} is used to show the impact of different choices of hyperparameters for the heavy-tailed priors, both in conflicting and non-conflicting situations. In \autoref{fig:vary_mu and_lambda}, we observe what has been explained before. Firstly, a Student prior resolves a conflict due to a prior location significantly different than that of the likelihood slower than a LPTN prior, i.e.\ the convergence towards the limiting posterior distribution is slower as $\mu_2 \rightarrow \infty$. Here, the limiting posterior resulting from a Student prior is not much different to that resulting from a LPTN prior; in both cases, the distribution of $\beta_2 \mid \sigma, \mathbf{y}$ is the same, but the distribution of $\sigma \mid \mathbf{y}$ is such that $\sigma^2$ follows an inverse-gamma and in the former case the shape and scale parameters are $(n - \gamma - 2) / 2 = 47$ and $n / 2$, respectively, whereas in the latter case, they are $(n - 2) / 2 = 49$ and $n / 2$, respectively. Similar arguments explains why, if we set $g_2$ to a CTN distribution and set the prior density of $\sigma$ such that it is proportional to $\sigma \pi(\sigma)$ to correct for the trace asymptotically left by a CTN prior distribution, we obtain essentially the same estimates and standard deviations as if we did not correct for this trace and instead set the prior on $\sigma$ to $\pi(\, \cdot \,)$ (the lines are on top of each other in \autoref{fig:vary_mu and_lambda}). In practice, one does the latter. Note that when we correct for that trace, we do it regardless of the values of $\mu_2$ and $\lambda_2$, and therefore, for some values, we should not correct because the situations are non-conflicting. The correction is needed in the asymptotic regime, which is something theoretical, explaining why we did not discriminate.

 In \autoref{fig:vary_mu and_lambda}, we also observe that using a Student or a LPTN prior is ineffective at resolving a conflict due to extremely small scalings (represented by $\lambda_2 \rightarrow \infty$), contrarily to using a CTN prior. A last point to note in \autoref{fig:vary_mu and_lambda} is that, because the LPTN distribution is the most similar to the standard normal distribution among the heavy-tailed distributions presented, using a LPTN prior translates into the closest results with those produced by using a normal one when there is no conflict, but also into the largest impact in the ``gray'' area, i.e.\ in between no conflict and clear conflict.

 In \autoref{fig:vary_mu_mean_sd_hyper}, we observe that increasing the level of similarity between an heavy-tailed prior and a normal prior (controlled through $\gamma$, $\rho$ and $\varrho$ for the Student, LPTN and CTN prior distributions, respectively) increases the threshold at which the Bayesian model starts to detect that the prior is conflicting (i.e.\ the point beyond which the impact starts to decrease) and thus increases the impact on the posterior distribution and estimate at this threshold. Our simulation study suggests that $\gamma =4$, $\rho = 0.95$ and $\varrho = 0.98$ offer a good balance between great similarity with the standard normal (and thus great similarity in between the posterior distributions in the absence of conflict) and great capacity at detecting and resolving a conflict (due to a prior location significantly different than that of the likelihood for Student and LPTN priors). The impact of hyperparameters when there is a conflict due to small scalings is not shown because showing it is relevant only for CTN prior distributions, and the impact is similar as when the conflict is due to a prior location significantly different than that of the likelihood.

\normalsize

\normalsize

\section{Conclusion}\label{sec:conclusion}

In this paper, we characterized the impact of using heavy-tailed alternatives to normal prior distributions for regression coefficients. This was achieved through a theoretical analysis under an asymptotic regime for which a conflicting situation becomes extreme and a simulation study, in Sections \ref{sec:results} and \ref{sec:simulation}, respectively. The heavy-tailed alternatives are Student, LPTN and CTN prior distributions. With the results presented in hand, one is well equipped to decide which prior distributions to use for a Bayesian regression analysis. In summary, normal prior distributions can be used when one is confident that they will not be in conflict with the data to collect; otherwise, heavy-tailed alternatives should be employed. All heavy-tailed alternatives can be used in a situation of a potential conflict due to a prior location significantly different than that of the likelihood function. Using Student and CTN prior distributions has an impact on the posterior variability of all coefficients asymptotically as the conflict becomes extreme; the variability increases when using Student prior distributions, while it decreases when using CTN prior distributions. The impact is however small when the sample size is large relatively to the number of conflicting prior densities. Note that this is however only true for Student priors with small degrees of freedom. When the priors on the regression coefficients are such that one is exposed to potential conflicts due to prior scalings, the heavy-tailed alternative that is recommended is the CTN distribution.

The theoretical analysis performed in \autoref{sec:results} was under the framework of linear regression. While there is no reason to believe that the results do not hold under other regression frameworks, like with GLMs, it would be interesting to prove similar results under such frameworks to have a confirmation and to have access to precise statements describing the conditions under which the results hold.

\bibliographystyle{rss}
\bibliography{reference}

\section{Acknowledgements}

The author acknowledges support from NSERC (Natural Sciences and Engineering Research Council of Canada) and FRQNT (Le Fonds de recherche du Qu\'{e}bec -- Nature et technologies). Also, the author thanks two anonymous referees and an associate editor for helpful suggestions that led to an improved manuscript.

\appendix

\section{Proofs}\label{sec_proofs}

The proof of Proposition 1 can be found in \cite{desgagne2015robustness}. In this section, we present the proofs of Theorems 1 and 3. The proof of Theorem 2 is an adaptation of that of Theorem 1 where we consider that $\text{C}_{\text{d}} = \varnothing$ and is thus omitted.

\begin{proof}[Proof of Theorem 1]
First, we prove that
 \begin{align*}
  \frac{m_\omega(\mathbf{y})}{\prod_{j \in \text{C}_{\text{b}}} g_j(\mu_j) \prod_{j \in \text{C}_{\text{d}}} \lambda_j \, g_j(\kappa)} \rightarrow \overline{m}(\mathbf{y}),
 \end{align*}
 with $m_\omega(\mathbf{y}) / [\prod_{j \in \text{C}_{\text{b}}} g_j(\mu_j) \prod_{j \in \text{C}_{\text{d}}} \lambda_j \, g_j(\kappa)] < \infty$ and $\overline{m}(\mathbf{y})< \infty$. Next, we prove that the posterior density converges pointwise. Finally, we prove the convergence of the posterior distribution.

 Assume for now that $m_\omega(\mathbf{y}) < \infty$ and $\overline{m}(\mathbf{y}) < \infty$; this will be shown later. We first observe that
 \begin{align*}
  \frac{m_\omega(\mathbf{y})}{\prod_{j \in \text{C}_{\text{b}}} g_j(\mu_j) \prod_{j \in \text{C}_{\text{d}}} \lambda_j \, g_j(\kappa) \, \overline{m}(\mathbf{y})} &= \frac{m_\omega(\mathbf{y})}{\prod_{j \in \text{C}_{\text{b}}} g_j(\mu_j) \prod_{j \in \text{C}_{\text{d}}} \lambda_j \, g_j(\kappa) \, \overline{m}(\mathbf{y})} \int_{\re^p}\int_0^\infty \pi_\omega(\boldsymbol\beta, \sigma \mid \mathbf{y}) \, \d\sigma \, \d\boldsymbol\beta \cr
  &= \int_{\re^p}\int_0^\infty \overline{\pi}(\boldsymbol\beta, \sigma \mid \mathbf{y}) \left[\prod_{j \in \text{C}_{\text{b}}} \frac{\pi_{j, \omega}(\beta_j \mid \sigma)}{g_j(\mu_j)}\prod_{j \in \text{C}_{\text{d}}} \frac{\pi_{j, \omega}(\beta_j \mid \sigma)}{(\lambda_j / \sigma)  g_j(\kappa)} \right] \, \d\sigma \, \d\boldsymbol\beta,
 \end{align*}
 where we used that
 \begin{align*}
 \pi_\omega(\boldsymbol\beta, \sigma \mid \mathbf{y}) = \pi(\sigma) \prod_{j = 1}^p \frac{\lambda_j}{\sigma} g_j\left(\frac{\lambda_j}{\sigma} (\beta_j - \mu_j)\right) \left[\prod_{i = 1}^n (1 / \sigma) f((y_i - \mathbf{x}_i^T \boldsymbol\beta) / \sigma)\right] \Bigg/ m_\omega(\mathbf{y}),
 \end{align*}
 and
  \begin{align*}
  \overline{\pi}(\boldsymbol\beta, \sigma \mid \mathbf{y}) := \pi(\sigma) \prod_{j \in \text{C}^\mathsf{c}} \pi_j(\beta_j \mid \sigma) \left[\prod_{i = 1}^n (1 / \sigma) f((y_i - \mathbf{x}_i^T \boldsymbol\beta) / \sigma)\right] \Bigg/ \overline{m}(\mathbf{y}).
   \end{align*}
 We show that the last integral converges towards 1 as $\omega \rightarrow \infty$. If we use Lebesgue's dominated convergence theorem to interchange the limit $\omega \rightarrow \infty$ and the integral, we have
 \begin{align*}
    &\lim_{\omega \rightarrow \infty} \int_{\re^p}\int_0^\infty \overline{\pi}(\boldsymbol\beta, \sigma \mid \mathbf{y}) \left[\prod_{j \in \text{C}_{\text{b}}} \frac{\pi_{j, \omega}(\beta_j \mid \sigma)}{g_j(\mu_j)}\prod_{j \in \text{C}_{\text{d}}} \frac{\pi_{j, \omega}(\beta_j \mid \sigma)}{(\lambda_j / \sigma)  g_j(\kappa)} \right] \, \d\sigma \, \d\boldsymbol\beta \cr
    & \qquad = \int_{\re^p}\int_0^\infty \overline{\pi}(\boldsymbol\beta, \sigma \mid \mathbf{y}) \lim_{\omega \rightarrow \infty} \left[\prod_{j \in \text{C}_{\text{b}}} \frac{\pi_{j, \omega}(\beta_j \mid \sigma)}{g_j(\mu_j)}\prod_{j \in \text{C}_{\text{d}}} \frac{\pi_{j, \omega}(\beta_j \mid \sigma)}{(\lambda_j / \sigma)  g_j(\kappa)} \right] \, \d\sigma \, \d\boldsymbol\beta \cr
    & \qquad = \int_{\re^p}\int_0^\infty \overline{\pi}(\boldsymbol\beta, \sigma \mid \mathbf{y}) \, \d\sigma \, \d\boldsymbol\beta = 1,
 \end{align*}
 by Proposition 1, the definition of CTN density (see, e.g., (4) in the manuscript) and using that the limiting posterior density is proper (which will be proven later). Note that pointwise convergence is sufficient, for any value of $\boldsymbol\beta \in \re^p$ and $\sigma > 0$, once the limit is inside the integral. Note also that we do not have convergence on the set $\bigcup_j \{\beta_j: j \in \text{C}_{\text{d}} \text{ and } \beta_j = \mu_j\}$, but this set has null measure so it does not affect the integral and the limit.

 In order to use Lebesgue's dominated convergence theorem, we need to prove that the integrand is bounded above by an integrable function of $\boldsymbol\beta$ and $\sigma$ that does not depend on $\omega$. Under the framework described in Section 3.1, we have that all $g_j$ are bounded and strictly positive, therefore we can choose the constant $M$ such that, for all $j \in \text{C}_{\text{d}}$,
 \[
  \frac{\pi_{j, \omega}(\beta_j \mid \sigma)}{(\lambda_j / \sigma)  g_j(\kappa)} = \frac{g_j\left(\frac{\lambda_j}{\sigma} (\beta_j - \mu_j) \right)}{g_j(\kappa)} \leq M,
 \]
 implying that
 \begin{align*}
  & \overline{\pi}(\boldsymbol\beta, \sigma \mid \mathbf{y}) \left[\prod_{j \in \text{C}_{\text{b}}} \frac{\pi_{j, \omega}(\beta_j \mid \sigma)}{g_j(\mu_j)}\prod_{j \in \text{C}_{\text{d}}} \frac{\pi_{j, \omega}(\beta_j \mid \sigma)}{(\lambda_j / \sigma)g_j(\kappa)} \right] \leq M^{|\text{C}_{\text{d}}|} \, \overline{\pi}(\boldsymbol\beta, \sigma \mid \mathbf{y}) \left[\prod_{j \in \text{C}_{\text{b}}} \frac{\pi_{j, \omega}(\beta_j \mid \sigma)}{g_j(\mu_j)}\right].
 \end{align*}
 We have that
 \begin{align*}
 \overline{\pi}(\boldsymbol\beta, \sigma \mid \mathbf{y}) \left[\prod_{j \in \text{C}_{\text{b}}} \frac{\pi_{j, \omega}(\beta_j \mid \sigma)}{g_j(\mu_j)}\right] &= [\overline{m}(\mathbf{y})]^{-1} \, \pi(\sigma) \left[\prod_{j \in \text{C}^\mathsf{c}} \pi_j(\beta_j \mid \sigma)\right] \left[\prod_{i = 1}^n \frac{1}{\sigma} f\left(\frac{y_i - \mathbf{x}_i^T \boldsymbol\beta}{\sigma}\right)\right]  \left[\prod_{j \in \text{C}_{\text{b}}} \frac{\pi_{j, \omega}(\beta_j \mid \sigma)}{g_j(\mu_j)}\right] \cr
 &\leq M^{n + 2 |\text{C}^\mathsf{c}| + |\text{C}_{\text{b}}|} [\overline{m}(\mathbf{y})]^{-1} \, \pi(\sigma) \left[\prod_{j \in \text{C}^\mathsf{c}} \frac{1}{\sigma} g_{\text{LPTN}}\left(\frac{\lambda_j}{\sigma} (\beta_j - \mu_j) \right)\right]   \cr
 &\qquad \times \left[\prod_{i = 1}^n \frac{1}{\sigma} g_{\text{LPTN}}\left(\frac{y_i - \mathbf{x}_i^T \boldsymbol\beta}{\sigma}\right)\right] \left[\prod_{j \in \text{C}_{\text{b}}} \frac{\frac{1}{\sigma} g_{\text{LPTN}}\left(\frac{\lambda_j}{\sigma} (\beta_j - \mu_j) \right)}{g_{\text{LPTN}}(\mu_j)}\right],
 \end{align*}
 using that, for $j \in \text{C}^\mathsf{c}$,
 \[
  \pi_j(\beta_j \mid \sigma) = \frac{\lambda_j}{\sigma} g_j\left(\frac{\lambda_j}{\sigma} (\beta_j - \mu_j) \right) \leq M^2 \frac{1}{\sigma} g_{\text{LPTN}}\left(\frac{\lambda_j}{\sigma} (\beta_j - \mu_j) \right),
 \]
 because $g_j / g_{\text{LPTN}} \leq M$ and considering that we can choose the constant $M$ such that $\lambda_j = c_j \leq M$,
 \[
  f\left(\frac{y_i - \mathbf{x}_i^T \boldsymbol\beta}{\sigma}\right) \leq M g_{\text{LPTN}}\left(\frac{y_i - \mathbf{x}_i^T \boldsymbol\beta}{\sigma}\right),
 \]
 and finally, that for $j \in \text{C}_{\text{b}}$, $g_j = g_{\text{LPTN}}$ and considering that we can choose the constant $M$ such that $\lambda_j = c_j \leq M$.

Consequently,
 \begin{align*}
   &\overline{\pi}(\boldsymbol\beta, \sigma \mid \mathbf{y}) \left[\prod_{j \in \text{C}_{\text{b}}} \frac{\pi_{j, \omega}(\beta_j \mid \sigma)}{g_j(\mu_j)}\right] \cr
  &\qquad \leq  \frac{M^{n + 2 |\text{C}^\mathsf{c}| + |\text{C}_{\text{b}}|}[\overline{m}(\mathbf{y})]^{-1}}{[\overline{m}_{\text{LPTN}}(\mathbf{y})]^{-1}} \, \overline{\pi}_{\text{LPTN}}(\boldsymbol\beta, \sigma \mid \mathbf{y})  \left[\prod_{j \in \text{C}_{\text{b}}} \frac{\frac{1}{\sigma} g_{\text{LPTN}}\left(\frac{\lambda_j}{\sigma} (\beta_j - \mu_j) \right)}{g_{\text{LPTN}}(\mu_j)}\right],
 \end{align*}
 where
 \[
  \overline{\pi}_{\text{LPTN}}(\boldsymbol\beta, \sigma \mid \mathbf{y}) :=  [\overline{m}_{\text{LPTN}}(\mathbf{y})]^{-1} \,  \pi(\sigma) \left[\prod_{j \in \text{C}^\mathsf{c}} \frac{1}{\sigma} g_{\text{LPTN}}\left(\frac{\lambda_j}{\sigma} (\beta_j - \mu_j) \right)\right]  \left[\prod_{i = 1}^n \frac{1}{\sigma} g_{\text{LPTN}}\left(\frac{y_i - \mathbf{x}_i^T \boldsymbol\beta}{\sigma}\right)\right],
 \]
 with
 \[
  \overline{m}_{\text{LPTN}}(\mathbf{y}) := \int_{\re^p}\int_0^\infty \pi(\sigma)  \left[\prod_{j \in \text{C}^\mathsf{c}} \frac{1}{\sigma} g_{\text{LPTN}}\left(\frac{\lambda_j}{\sigma} (\beta_j - \mu_j) \right)\right]  \left[\prod_{i = 1}^n \frac{1}{\sigma} g_{\text{LPTN}}\left(\frac{y_i - \mathbf{x}_i^T \boldsymbol\beta}{\sigma}\right)\right] \, \d\sigma \, \d\boldsymbol\beta,
 \]
 assuming that $\overline{m}_{\text{LPTN}}(\mathbf{y}) < \infty$ (we prove this below). We can prove that
 \[
  \overline{\pi}_{\text{LPTN}}(\boldsymbol\beta, \sigma \mid \mathbf{y})  \left[\prod_{j \in \text{C}_{\text{b}}} \frac{\frac{1}{\sigma} g_{\text{LPTN}}\left(\frac{\lambda_j}{\sigma} (\beta_j - \mu_j) \right)}{g_{\text{LPTN}}(\mu_j)}\right]
 \]
 is bounded above by an integrable function of $\boldsymbol\beta$ and $\sigma$ that does not depend on $\omega$ in the same way that it is done in the proof of Theorem 2.1, Result (a), in \cite{gagnon2020supp} because the function above represents a special case of that in that proof. In \cite{gagnon2020}, the theoretical result is about the convergence of the posterior distribution as some $y_j \rightarrow \pm \infty$ in a context of robustness against outliers. It is considered that the joint prior on all parameters is non-conflicting and that it is bounded above by $\max(M, \sigma^{-1} M)$. To fit within the framework of \cite{gagnon2020}, we treat $\lambda_j \mu_j$ for $j \in \text{C}^\mathsf{c} \bigcup \text{C}_{\text{b}}$ as an observation from the dependent variable and $\lambda_j$ that multiplies $\beta_j$ for $j$ in the same set as a vector of covariates where the other covariates are all equal to 0. There is no problem with the fact that the first component of these vectors are not 1. The sample size in the framework of \cite{gagnon2020} thus corresponds to $n + |\text{C}^\mathsf{c}| + |\text{C}_{\text{b}}|$ here.

 What allows to exploit the proof in \cite{gagnon2020supp} is that the assumptions of Theorem 2.1 are verified. It is readily seen that the following allows to verify the assumptions:
\begin{itemize}
\itemsep 0mm

 \item the density of all ``observations'' (including the prior distributions with $j \in \text{C}^\mathsf{c} \bigcup \text{C}_{\text{b}}$) are LPTN;

 \item $\pi(\sigma) \leq \max(M, \sigma^{-1} M)$;

 \item $n + |\text{C}^\mathsf{c}| \geq 2p + 1 + |\text{C}_{\text{b}}|$.

\end{itemize}
This concludes the proof that
\begin{align*}
  \frac{m_\omega(\mathbf{y})}{\prod_{j \in \text{C}_{\text{b}}} g_j(\mu_j) \prod_{j \in \text{C}_{\text{d}}} \lambda_j \, g_j(\kappa)} \rightarrow \overline{m}(\mathbf{y}),
 \end{align*}
 assuming that $\overline{m}_{\text{LPTN}}(\mathbf{y}) < \infty$, $\overline{m}(\mathbf{y})< \infty$ and  $m_\omega(\mathbf{y}) < \infty$ for all $\omega$.

We now show that under the conditions above, $\overline{m}_{\text{LPTN}}(\mathbf{y}) < \infty$, which will be seen to imply that $\overline{m}(\mathbf{y})< \infty$ and  $m_\omega(\mathbf{y}) / [\prod_{j \in \text{C}_{\text{b}}} g_j(\mu_j) \prod_{j \in \text{C}_{\text{d}}} \lambda_j \, g_j(\kappa)] < \infty$ (which in turn implies that $m_\omega(\mathbf{y}) < \infty$ for all $\omega$). We proceed as follows: first we show that $\overline{m}(\mathbf{y})$ is bounded above by a constant times $\overline{m}_{\text{LPTN}}(\mathbf{y})$, next we show that $\overline{m}_{\text{LPTN}}(\mathbf{y}) < \infty$. This will allow to conclude that $m_\omega(\mathbf{y}) / [\prod_{j \in \text{C}_{\text{b}}} g_j(\mu_j) \prod_{j \in \text{C}_{\text{d}}} \lambda_j \, g_j(\kappa)] < \infty$ because we will have shown that
 \[
 \frac{m_\omega(\mathbf{y})}{\prod_{j \in \text{C}_{\text{b}}} g_j(\mu_j) \prod_{j \in \text{C}_{\text{d}}} \lambda_j \, g_j(\kappa) \, \overline{m}(\mathbf{y})}
 \]
 is bounded above by a constant times an integral of an integrable function.

 We have
 \begin{align*}
    \overline{m}(\mathbf{y}) &=  \int_{\re^p} \int_0^\infty \pi(\sigma) \left[\prod_{j \in \text{C}^\mathsf{c}} \frac{\lambda_j}{\sigma} g_{j}\left(\frac{\lambda_j}{\sigma} (\beta_j - \mu_j) \right)\right] \left[\prod_{i=1}^n \frac{1}{\sigma} f\left(\frac{y_i - \mathbf{x}_i^T \boldsymbol\beta}{\sigma}\right) \right] \, \d\sigma \, \d\boldsymbol\beta \cr
    &\leq M^{n + 2|\text{C}^\mathsf{c}|} \int_{\re^p} \int_0^\infty \pi(\sigma) \left[\prod_{j \in \text{C}^\mathsf{c}} \frac{1}{\sigma} g_{\text{LPTN}}\left(\frac{\lambda_j}{\sigma} (\beta_j - \mu_j) \right)\right]  \left[\prod_{i = 1}^n \frac{1}{\sigma} g_{\text{LRVD}}\left(\frac{y_i - \mathbf{x}_i^T \boldsymbol\beta}{\sigma}\right)\right]\, \d\sigma \, \d\boldsymbol\beta \propto \overline{m}_{\text{LRVD}}(\mathbf{y}),
 \end{align*}
 using the same arguments as above. Proving that $\overline{m}_{\text{LPTN}}(\mathbf{y})$ is finite is done in the same way as in the proof of Proposition 2.1 in \cite{gagnon2020supp} because the integrand above represents a special case of that in \cite{gagnon2020supp}. As previously, what allows to exploit the proof in \cite{gagnon2020supp} is that the assumptions of Proposition 2.1 are verified. It is readily seen that the following allows to verify the assumptions:
\begin{itemize}
\itemsep 0mm

 \item the density of all ``observations'' (including the prior distributions with $j \in \text{C}^\mathsf{c} \bigcup \text{C}_{\text{b}}$) are LPTN;

 \item $\pi(\sigma) \leq \max(M, \sigma^{-1} M)$;

 \item $n + |\text{C}^\mathsf{c}| \geq 2p + 1 + |\text{C}_{\text{b}}|$, implying that $n + |\text{C}^\mathsf{c}| \geq p + 1$.

\end{itemize}
This concludes the proof that $\overline{m}_{\text{LPTN}}(\mathbf{y}) < \infty$, $\overline{m}(\mathbf{y})< \infty$ and $m_\omega(\mathbf{y}) < \infty$ for all $\omega$.

We now prove that the posterior density converges pointwise. We have that
    \begin{align*}
     \pi_\omega(\boldsymbol\beta, \sigma \mid \mathbf{y}) &= \overline{\pi}(\boldsymbol\beta, \sigma \mid \mathbf{y}) \frac{\overline{m}(\mathbf{y})}{m_\omega(\mathbf{y})} \sigma^{|\text{C}_{\text{d}}|} \left[\prod_{j \in \text{C}} \pi_{j, \omega}(\beta_j \mid \sigma) \right]  \cr
     &= \overline{\pi}(\boldsymbol\beta, \sigma \mid \mathbf{y}) \frac{\overline{m}(\mathbf{y})}{m_\omega(\mathbf{y})}\prod_{j \in \text{C}_{\text{b}}} g_j(\mu_j) \prod_{j \in \text{C}_{\text{d}}} \lambda_j \, g_j(\kappa) \left[\prod_{j \in \text{C}_{\text{b}}} \frac{\pi_{j, \omega}(\beta_j \mid \sigma)}{g_j(\mu_j)}\prod_{j \in \text{C}_{\text{d}}} \frac{\pi_{j, \omega}(\beta_j \mid \sigma)}{(\lambda_j / \sigma) g_j(\kappa)} \right],
    \end{align*}
    and
    \[
     \frac{\overline{m}(\mathbf{y})}{m_\omega(\mathbf{y})}\prod_{j \in \text{C}_{\text{b}}} g_j(\mu_j) \prod_{j \in \text{C}_{\text{d}}} \lambda_j \, g_j(\kappa) \left[\prod_{j \in \text{C}_{\text{b}}} \frac{\pi_{j, \omega}(\beta_j \mid \sigma)}{g_j(\mu_j)}\prod_{j \in \text{C}_{\text{d}}} \frac{\pi_{j, \omega}(\beta_j \mid \sigma)}{(\lambda_j / \sigma) g_j(\kappa)} \right] \rightarrow 1,
    \]
    for any $\boldsymbol\beta \in \re^p, \sigma > 0$ using Proposition 1, the definition of CTN density (see, e.g., (4) in the manuscript) and the asymptotic behaviour of the marginal density, except on $\bigcup_j \{\beta_j: j \in \text{C}_{\text{d}} \text{ and } \beta_j = \mu_j\}$. On this set,
    \[
     \frac{\overline{m}(\mathbf{y})}{m_\omega(\mathbf{y})}\prod_{j \in \text{C}_{\text{b}}} g_j(\mu_j) \prod_{j \in \text{C}_{\text{d}}} \lambda_j \, g_j(\kappa) \left[\prod_{j \in \text{C}_{\text{b}}} \frac{\pi_{j, \omega}(\beta_j \mid \sigma)}{g_j(\mu_j)}\right] \rightarrow 1
    \]
    but
     \[
     \frac{\pi_{j, \omega}(\beta_j \mid \sigma)}{(\lambda_j / \sigma) g_j(\kappa)} = \frac{g_j(0)}{g_j(\kappa)},
    \]
    for some $ j$. Therefore, the limiting value for $\pi_\omega(\boldsymbol\beta, \sigma \mid \mathbf{y})$ on this set is $\overline{\pi}(\boldsymbol\beta, \sigma \mid \mathbf{y})$ times a factor. This concludes the proof that the posterior density converges pointwise, except on a set of null measure.

  Now that we know that the posterior density converges pointwise (except on a set of null measure), the convergence of the posterior distribution follows directly using Scheffé's theorem (see \cite{scheffe1947useful}).
\end{proof}

\begin{proof}[Proof of Theorem 3]
 We proceed as in the previous proof:
 \begin{align*}
  \frac{m_\omega(\mathbf{y})}{\prod_{j \in \text{C}} \lambda_j \, g_j(\kappa) \, \overline{m}(\mathbf{y})} &= \frac{m_\omega(\mathbf{y})}{\prod_{j \in \text{C}} \lambda_j \, g_j(\kappa) \, \overline{m}(\mathbf{y})} \int_{\re^p}\int_0^\infty \pi_\omega(\boldsymbol\beta, \sigma \mid \mathbf{y}) \, \d\sigma \, \d\boldsymbol\beta \cr
  &= \int_{\re^p}\int_0^\infty \overline{\pi}(\boldsymbol\beta, \sigma \mid \mathbf{y}) \left[\prod_{j \in \text{C}} \frac{g_{\text{CTN}}\left(\frac{\lambda_j}{\sigma} (\beta_j - \mu_j) \right)}{g_{\text{CTN}}(\kappa)} \right] \, \d\sigma \, \d\boldsymbol\beta.
 \end{align*}
 We show that the last integral converges towards 1 as $\omega \rightarrow \infty$. If we use Lebesgue's dominated convergence theorem to interchange the limit $\omega \rightarrow \infty$ and the integral, we have
 \begin{align*}
    &\lim_{\omega \rightarrow \infty} \int_{\re^p}\int_0^\infty \overline{\pi}(\boldsymbol\beta, \sigma \mid \mathbf{y}) \left[\prod_{j \in \text{C}} \frac{g_{\text{CTN}}\left(\frac{\lambda_j}{\sigma} (\beta_j - \mu_j) \right)}{g_{\text{CTN}}(\kappa)} \right] \, \d\sigma \, \d\boldsymbol\beta \cr
    & \qquad = \int_{\re^p}\int_0^\infty \overline{\pi}(\boldsymbol\beta, \sigma \mid \mathbf{y}) \lim_{\omega \rightarrow \infty} \left[\prod_{j \in \text{C}} \frac{g_{\text{CTN}}\left(\frac{\lambda_j}{\sigma} (\beta_j - \mu_j) \right)}{g_{\text{CTN}}(\kappa)} \right] \, \d\sigma \, \d\boldsymbol\beta \cr
    & \qquad = \int_{\re^p}\int_0^\infty \overline{\pi}(\boldsymbol\beta, \sigma \mid \mathbf{y}) \, \d\sigma \, \d\boldsymbol\beta = 1,
 \end{align*}
 using the definition of CTN density (see, e.g., (4) in the manuscript) and using that the limiting posterior distribution is proper (which will be proven later). Note that pointwise convergence is sufficient, for any value of $\boldsymbol\beta \in \re^p$ and $\sigma > 0$, once the limit is inside the integral. Note also that we do not have convergence on the set $\bigcup_j \{\beta_j: j \in \text{C}_{\text{d}} \text{ and } \beta_j = \mu_j\}$, but this set has null measure so it does not affect the integral and the limit.

  In order to use Lebesgue's dominated convergence theorem, we need to prove that the integrand is bounded above by an integrable function of $\boldsymbol\beta$ and $\sigma$ that does not depend on $\omega$. The proof under the framework of Theorem 3 is easier than that under the framework of Theorem 1 and does not rely on the proof of Theorem 2.1 in \cite{gagnon2020supp}. Under the framework described in Section 3.1 and by the definition of CTN density, we have
  \begin{align*}
   \overline{\pi}(\boldsymbol\beta, \sigma \mid \mathbf{y}) \left[\prod_{j \in \text{C}} \frac{g_{\text{CTN}}\left(\frac{\lambda_j}{\sigma} (\beta_j - \mu_j) \right)}{g_{\text{CTN}}(\kappa)} \right]  &\leq  \overline{\pi}(\boldsymbol\beta, \sigma \mid \mathbf{y}) \, M^{|\text{C}|},
  \end{align*}
  because $g_{\text{CTN}}$ is bounded from above and it is strictly positive. There thus only remains to prove that $\overline{\pi}(\, \cdot \,, \cdot \mid \mathbf{y})$ is proper, which will imply that $m_\omega(\mathbf{y}) < \infty$ for all $\omega$. Indeed, to prove that $\overline{\pi}(\, \cdot \,, \cdot \mid \mathbf{y})$ is proper, we prove that $\overline{m}(\mathbf{y}) < \infty$; we will thus have shown that
 \[
 \frac{m_\omega(\mathbf{y})}{\prod_{j \in \text{C}} \lambda_j \, g_j(\kappa) \, \overline{m}(\mathbf{y})}
 \]
 is bounded above by a constant times an integral of an integrable function and that $\overline{m}(\mathbf{y}) < \infty$.

  We prove that $\overline{m}(\mathbf{y}) < \infty$ similarly as in the proof of Proposition 2.1 in \cite{gagnon2020supp}. We have that
 \begin{align*}
    \overline{m}(\mathbf{y}) &=   \int_{\re^p} \int_0^\infty \sigma^{-|\text{C}|} \, \pi(\sigma) \prod_{j \in \text{C}^\mathsf{c}} \frac{\lambda_j}{\sigma} \,g_{\text{CTN}}\left(\frac{\lambda_j}{\sigma} (\beta_j - \mu_j) \right) \prod_{i = 1}^n \frac{1}{\sigma} f\left(\frac{y_i - \mathbf{x}_i^T \boldsymbol\beta}{\sigma}\right) \, \d\sigma \, \d\boldsymbol\beta \cr
    &\leq M^{|\text{C}^\mathsf{c}|} \int_{\re^p} \int_0^\infty \sigma^{-p} \, \pi(\sigma) \prod_{i = 1}^n \frac{1}{\sigma} f\left(\frac{y_i - \mathbf{x}_i^T \boldsymbol\beta}{\sigma}\right) \, \d\sigma \, \d\boldsymbol\beta,
 \end{align*}
 because we can choose $M$ such that
 \[
  \lambda_j \, g_{\text{CTN}}\left(\frac{\lambda_j}{\sigma} (\beta_j - \mu_j) \right) \leq M
 \]
 for all $j \in \text{C}^\mathsf{c}$; recall that for $j \in \text{C}^\mathsf{c}$, $\lambda_j = c_j$.

 We now prove that the integral above is finite. We first show that the function is integrable on an area where the ratio $1/\sigma$ is bounded. More precisely, we consider $\boldsymbol\beta\in\re^p$ and  $\delta M^{-1}\le \sigma<\infty$, where $\delta$ is a positive constant that can be chosen as small as we want (upper bounds are provided in the proof). We next show that the function is integrable on the complement set where the ratio $1/\sigma$ approaches infinity, that is $0< \sigma<\delta M^{-1}$. We have
\begin{align*}
&\int_{\delta M^{-1}}^{\infty}\int_{\re^p}\sigma^{-p} \, \pi(\sigma) \prod_{i = 1}^n \frac{1}{\sigma} f\left(\frac{y_i - \mathbf{x}_i^T \boldsymbol\beta}{\sigma}\right) \, \d\sigma \, \d\boldsymbol\beta \\
&\quad\za{\le} (\delta^{-1} M)^{n} M^{n - p} \int_{\delta M^{-1}}^{\infty} \sigma^{-p} \, \pi(\sigma) \int_{\re^p}\prod_{i=1}^p \frac{1}{\sigma} f\left(\frac{y_i - \mathbf{x}_i^T \boldsymbol\beta}{\sigma}\right) \,\d\boldsymbol\beta \,\d\sigma\\
&\quad\zb{\leq}  (\delta^{-1} M)^{n} M^{n - p} \abs{\text{det}\left(\begin{array}{c}\mathbf{x}_1^T \cr \vdots \cr \mathbf{x}_p^T\end{array}\right)}^{-1}\int_{0}^{\infty} \sigma^{-p} \, \pi(\sigma)  \,\d\sigma \prod_{i=1}^p \int_{-\infty}^{\infty} f(u_i) \,\d u_i \zc{<} \infty.
\end{align*}
In Step $a$, we bound each of $n - p$ densities $f$ by $M$, requiring that $n\geq p$, and $p + n - p = n$ times $\sigma^{-1}$ using $\sigma^{-1} \leq \delta^{-1} M$. In Step $b$, we use that $[\delta M^{-1}, \infty) \subset \re$ and the change of variables $u_i=(y_i-\mathbf{x}_i^T\boldsymbol\beta)/\sigma$ for $i=1,\ldots,p$. The determinant is non-null because all explanatory variables are continuous. In Step $c$, we use that $\int_{0}^{\infty} \sigma^{-p} \, \pi(\sigma)  \,\d\sigma < \infty$ and $f$ is a proper density.

We now show that the integral is finite on $\boldsymbol\beta\in\re^p$ and $0<\sigma<\delta M^{-1}$. On this area, the ratio $(1/\sigma)$ approaches infinity. We have to carefully analyse the sub-areas where the terms $y_i-\mathbf{x}_i^T\boldsymbol\beta$ are close to 0 in order to deal with the $0/0$ form of the ratios $(y_i-\mathbf{x}_i^T \boldsymbol\beta)/\sigma$. To achieve this, we split the domain of $\boldsymbol\beta$ as follows:
 \begin{align}\label{eqn_R_p}
  \re^p&=\left[\bigcap_{i_1=1}^n \mathcal{R}_{i_1}^\mathsf{c}\right] \bigcup \left[\bigcup_{i_1=1}^n\left(\mathcal{R}_{i_1}\bigcap\left(\bigcap_{i_2=1 (i_2\neq i_1)}^n \mathcal{R}_{i_2}^\mathsf{c}\right)\right)\right] \bigcup\left[\bigcup_{i_1,i_2=1 (i_1\neq i_2)}^n  \left(\mathcal{R}_{i_1}\bigcap \mathcal{R}_{i_2}\bigcap\left(\bigcap_{i_3=1 (i_3\neq i_1,i_2)}^n\mathcal{R}_{i_3}^\mathsf{c} \right)\right)\right] \cr
  &\qquad \bigcup \cdots \bigcup \left[\bigcup_{i_1,i_2,\ldots,i_p=1 (i_{j}\neq i_s\,\forall i_j,i_s\text{ s.t. }j\neq s)}^n  \left(\mathcal{R}_{i_1}\bigcap \mathcal{R}_{i_2}\bigcap \ldots\bigcap \mathcal{R}_{i_p}\right)\right],
 \end{align}
 where $\mathcal{R}_{i}:=\{\boldsymbol\beta: |y_i-\mathbf{x}_i^T \boldsymbol\beta|<\delta \}, i\in\{1,\ldots,n\}$. The set $\mathcal{R}_i$ represents the hyperplanes characterized by the different values of $\boldsymbol\beta$ that satisfy $|y_i-\mathbf{x}_i^T \boldsymbol\beta|<\delta$. In other words, it represents the hyperplanes passing near the point $(\mathbf{x}_i,y_i)$, and more precisely, at a vertical distance of less than $\delta$. The set $\bigcap_{i_1=1}^n \mathcal{R}_{i_1}^\mathsf{c}$ is therefore comprised of the hyperplanes that are not passing close to any point. The set $\bigcup_{i_1=1}^n(\mathcal{R}_{i_1}\bigcap(\bigcap_{i_2=1 (i_2\neq i_1)}^n \mathcal{R}_{i_2}^\mathsf{c}))$ represents the hyperplanes passing near one (and only one) point, and so on.

 We choose $\delta$ small enough to ensure that $\mathcal{R}_{i_1}\bigcap \mathcal{R}_{i_2}\bigcap \ldots\bigcap \mathcal{R}_{i_p}\bigcap \mathcal{R}_{i_{p+1}}=\varnothing$ when $i_1,\ldots,i_{p+1}$ are all different. This is possible because an hyperplane passes through no more than $p$ points (in the situation where all explanatory variables are continuous). This implies that in \eqref{eqn_R_p}
\begin{align*}
 &\bigcup_{i_1,i_2,\ldots,i_p=1 (i_{j}\neq i_s\,\forall i_j,i_s\text{ s.t. }j\neq s)}^n  \left(\mathcal{R}_{i_1}\bigcap \mathcal{R}_{i_2}\bigcap \ldots\bigcap \mathcal{R}_{i_p}\right) \cr
 &\qquad=\bigcup_{i_1,i_2,\ldots,i_p=1 (i_{j}\neq i_s\,\forall i_j,i_s\text{ s.t. }j\neq s)}^n  \left(\mathcal{R}_{i_1}\bigcap \mathcal{R}_{i_2}\bigcap \ldots\bigcap \mathcal{R}_{i_p} \bigcap \left(\bigcap_{i_{p+1}=1 (i_{p+1}\neq i_1,i_2,\ldots,i_p)}^n\mathcal{R}_{i_{p+1}}^\mathsf{c} \right)\right).
\end{align*}
Note that all sets $\mathcal{R}_{i_1}\bigcap \mathcal{R}_{i_2}\bigcap \ldots\bigcap \mathcal{R}_{i_p}$ are non-empty when $i_1,\ldots,i_{p}$ are all different, because all explanatory variables are continuous. Note also that $\mathcal{R}_{i_1}\bigcap(\bigcap_{i_2=1 (i_2\neq i_1)}^n \mathcal{R}_{i_2}^\mathsf{c})$ is non-empty for all $i_1$, and so on. Finally note that the decomposition of $\re^p$ in \eqref{eqn_R_p} is made of $\sum_{i=0}^p {{n} \choose {i}}$ mutually exclusive sets given by $\bigcap_{i_1=1}^n \mathcal{R}_{i_1}^\mathsf{c}$, $\mathcal{R}_{i_1}\bigcap(\bigcap_{i_2=1 (i_2\neq i_1)}^n \mathcal{R}_{i_2}^\mathsf{c}),i_1=1,\ldots,n$, and so on.

We thus consider that $0<\sigma<\delta M^{-1}$ and $\boldsymbol\beta$ belongs to one of the $\sum_{i=0}^p {{n} \choose {i}}$ mutually exclusive sets given in (\ref{eqn_R_p}). As explained above, the difficulty lies in dealing with the hyperplanes parametrized by $\boldsymbol\beta$ that are such that $|y_i-\mathbf{x}_i^T \boldsymbol\beta|<\delta$ for some points $(\mathbf{x}_i,y_i)$. The strategy is to use the terms $(1/\sigma)f((y_i-\mathbf{x}_i^T \boldsymbol\beta)/\sigma )$ associated to these points to integrate over $\boldsymbol\beta$ (requiring $n \geq p$), and to bound the other terms. Therefore, if $\boldsymbol\beta\in\mathcal{R}_{i_1}\bigcap \mathcal{R}_{i_2}\bigcap \ldots\bigcap \mathcal{R}_{i_p}$, we use the points $(\mathbf{x}_{i_1},y_{i_1}),(\mathbf{x}_{i_2},y_{i_2}), \ldots, (\mathbf{x}_{i_p},y_{i_p})$ to integrate over $\boldsymbol\beta$. If $\boldsymbol\beta\in\mathcal{R}_{i_1}\bigcap \mathcal{R}_{i_2}\bigcap \ldots\mathcal{R}_{i_{p-1}}\bigcap (\bigcap_{i_p=1 (i_p\neq i_1,\ldots,i_{p-1})}^n\mathcal{R}_{i_p}^\mathsf{c}))$, we use the points $(\mathbf{x}_{i_1},y_{i_1}),(\mathbf{x}_{i_2},y_{i_2}), \ldots, (\mathbf{x}_{i_{p-1}},y_{i_{p-1}})$, and any other point $(\mathbf{x}_{i_{p}},y_{i_{p}})$ (leading to a matrix with a non-null determinant) to integrate over $\boldsymbol\beta$, and so on. We have
 \begin{align*}
  \sigma^{-p} \, \pi(\sigma) \prod_{i = 1}^n \frac{1}{\sigma} f\left(\frac{y_i - \mathbf{x}_i^T \boldsymbol\beta}{\sigma}\right) &= \sigma^{-p} \, \pi(\sigma) \prod_{i \in \{i_1,\ldots,i_p\}}\frac{1}{\sigma} f\left(\frac{y_i - \mathbf{x}_i^T \boldsymbol\beta}{\sigma}\right) \prod_{i \notin \{i_1,\ldots,i_p\}} \frac{1}{\sigma} f\left(\frac{y_i - \mathbf{x}_i^T \boldsymbol\beta}{\sigma}\right) \cr
  &\za{\leq} \sigma^{-p}\, \pi(\sigma) \left[\frac{1}{\sigma} f\left(\frac{\delta}{\sigma} \right)\right]^{n - p} \prod_{i \in \{i_1,\ldots,i_p\}}\frac{1}{\sigma} f\left(\frac{y_i - \mathbf{x}_i^T \boldsymbol\beta}{\sigma}\right)  \cr
  &\zb{\leq} [M/\delta]^{n-p} \sigma^{-p}\, \pi(\sigma) \prod_{i \in \{i_1,\ldots,i_p\}}\frac{1}{\sigma} f\left(\frac{y_i - \mathbf{x}_i^T \boldsymbol\beta}{\sigma}\right).
 \end{align*}
In Step $a$, for all $i \notin \{i_1,\ldots,i_p\}$, we use that $f((y_i-\mathbf{x}_i^T \boldsymbol\beta)/\sigma) \leq f(\delta/\sigma)$ by the monotonicity of the tails of $f$ because $|y_i-\mathbf{x}_i^T\boldsymbol\beta|/\sigma\ge \delta/\sigma\ge \delta\delta^{-1} M=M$ (we can consider that above this value, the function is monotonic because $\xi$ defined by $z \mapsto zf(z)$ is monotonic). In Step $b$, we bound $n-p$ terms $(1/\sigma)f(\delta/\sigma)$ by $M/\delta$ (given that $\xi$ is continuous and monotonic above a constant, it is bounded).

 Finally, we bound the integral of the function above by
\begin{align*}
&\int_{0}^{\infty} \sigma^{-p} \, \pi(\sigma) \int_{\re^p} \prod_{i \in \{i_1,\ldots,i_p\}}\frac{1}{\sigma} f\left(\frac{y_i - \mathbf{x}_i^T \boldsymbol\beta}{\sigma}\right) \,\d\boldsymbol\beta\,\d\sigma
 = \abs{\text{det}\left(\begin{array}{c}\mathbf{x}_{i_1}^T \cr \vdots \cr \mathbf{x}_{i_p}^T\end{array}\right)}^{-1}\int_{0}^{\infty} \sigma^{-p} \, \pi(\sigma) \,\d\sigma <\infty,
\end{align*}
using the same change of variables as above $u_j=(y_{i_j}-\mathbf{x}_{i_j}^T\boldsymbol\beta)/\sigma$ for $j=1,\ldots,p$, and that $\int_{0}^{\infty} \sigma^{-p} \, \pi(\sigma) \,d\sigma < \infty$. This concludes the proof that
 \begin{align*}
  \frac{m_\omega(\mathbf{y})}{\prod_{j \in \text{C}} \lambda_j \, g_j(\kappa) \, \overline{m}(\mathbf{y})} &\rightarrow 1,
 \end{align*}
 and that $\overline{m}(\mathbf{y}) < \infty$ and $m_\omega(\mathbf{y}) < \infty$ for all $\omega$.

We now prove that the posterior density converges pointwise. We have that
    \begin{align*}
     \pi_\omega(\boldsymbol\beta, \sigma \mid \mathbf{y}) &= \overline{\pi}(\boldsymbol\beta, \sigma \mid \mathbf{y}) \, \frac{\prod_{j \in \text{C}} \lambda_j \, g_j(\kappa) \, \overline{m}(\mathbf{y})}{m_\omega(\mathbf{y})} \left[\prod_{j \in \text{C}} \frac{g_{\text{CTN}}\left(\frac{\lambda_j}{\sigma} (\beta_j - \mu_j) \right)}{g_{\text{CTN}}(\kappa)} \right],
    \end{align*}
    and
    \[
     \frac{\prod_{j \in \text{C}} \lambda_j \, g_j(\kappa) \, \overline{m}(\mathbf{y})}{m_\omega(\mathbf{y})} \left[\prod_{j \in \text{C}} \frac{g_{\text{CTN}}\left(\frac{\lambda_j}{\sigma} (\beta_j - \mu_j) \right)}{g_{\text{CTN}}(\kappa)} \right] \rightarrow 1,
    \]
    for any $\boldsymbol\beta \in \re^p, \sigma > 0$ using the definition of CTN densities (see, e.g., (4) in the manuscript) and the asymptotic behaviour of the marginal density, except on $\bigcup_j \{\beta_j: j \in \text{C}_{\text{d}} \text{ and } \beta_j = \mu_j\}$. On this set, the expression above converges to $g_{\text{CTN}}(0)/g_{\text{CTN}}(\kappa)$ at some power. Therefore, the limiting value for $\pi_\omega(\boldsymbol\beta, \sigma \mid \mathbf{y})$ on this set is $\overline{\pi}(\boldsymbol\beta, \sigma \mid \mathbf{y})$ times a factor. This concludes the proof that the posterior density converges pointwise, except on a set of null measure.

  Now that we know that the posterior density converges pointwise (except on a set of null measure), the convergence of the posterior distribution follows directly using Scheffé's theorem (see \cite{scheffe1947useful}).
 \end{proof}

\section{Details of the simulation study}\label{sec:simulation_supp}

We use a HMC algorithm to sample from the posterior distribution. We thus apply a transformation on $\sigma$ to make it a variable on the real line. The original target density is such that:
\[
 \pi(\beta_2, \sigma \mid \mathbf{y}) \propto \frac{1}{\sigma^{n}}\exp\left(-\frac{n}{2 \sigma^2} \right) \frac{1}{\sigma}\exp\left(-\frac{n}{2\sigma^2}\beta_2^2\right) \frac{\lambda_2 n^{1/2}}{\sigma}g_2\left(\frac{\lambda_2 n^{1/2}}{\sigma}(\beta_2 - \mu_2)\right).
\]
We define $\nu := \log \sigma$, and thus,
\[
 \pi(\beta_2, \nu \mid \mathbf{y}) \propto \frac{1}{\ee^{(n - 1) \nu}}\exp\left(-\frac{n}{2 \ee^{2 \nu}} \right) \frac{1}{\ee^{\nu}}\exp\left(-\frac{n}{2\ee^{2\nu}}\beta_2^2\right) \frac{\lambda_2 n^{1/2}}{\ee^{\nu}}g_2\left(\frac{\lambda_2 n^{1/2}}{\ee^{\nu}}(\beta_2 - \mu_2)\right).
\]
The log density is such that (if we forget about the constants):
\[
 \log \pi(\beta_2, \nu \mid \mathbf{y}) = -\nu(n + 1) -\frac{n}{2 \ee^{2 \nu}}-\frac{n}{2\ee^{2 \nu}}\beta_2^2 + \log g_2\left(\frac{\lambda_2 n^{1/2}}{\ee^{\nu}}(\beta_2 - \mu_2)\right).
\]
The gradient is such that:
\[
 \frac{\partial}{\partial \beta_2} \log \pi(\beta_2, \nu \mid \mathbf{y}) =-\frac{n}{\ee^{2 \nu}}\beta_2 + \frac{\partial}{\partial \beta_2} \log g_2\left(\frac{\lambda_2 n^{1/2}}{\ee^{\nu}}(\beta_2 - \mu_2)\right),
\]
\[
 \frac{\partial}{\partial \nu} \log \pi(\beta_2, \nu \mid \mathbf{y}) =-(n + 1) + \frac{n}{\ee^{2\nu}}(1 + \beta_2^2) + \frac{\partial}{\partial \nu} \log g_2\left(\frac{\lambda_2 n^{1/2}}{\ee^{\nu}}(\beta_2 - \mu_2)\right).
\]
We now derive the required expressions for each choice of $g_2$.

\textbf{Standard normal distribution.} In this case, we actually have access to closed-form expressions for the posterior mean and variance. We thus do not have to apply a transformation and do not need to sample from the posterior distribution. This follows from:
\begin{align*}
 \pi(\beta_2, \sigma \mid \mathbf{y}) &\propto \frac{1}{\sigma^{n}}\exp\left(-\frac{n}{2 \sigma^2} \right) \frac{1}{\sigma}\exp\left(-\frac{n}{2\sigma^2}\beta_2^2\right) \frac{1}{\sigma}\exp\left(-\frac{\lambda_2^2 n}{2\sigma^2}(\beta_2 - \mu_2)^2\right) \cr
 &= \frac{1}{\sigma^{n + 1}}\exp\left(-\frac{n}{2 \sigma^2}\left(1 + \frac{\mu_2^2 \lambda_2^2}{\lambda_2^2 + 1}\right)\right) \frac{1}{\sigma}\exp\left(-\frac{n (1 + \lambda_2^2)}{2\sigma^2}\left(\beta_2 - \mu_2 \frac{\lambda_2^2}{1 + \lambda_2^2} \right)^2\right).
\end{align*}
This implies that $\beta_2 \mid \sigma \sim \mathcal{N}(\mu_2 \lambda_2^2 / (1 + \lambda_2^2), \sigma_2^2 / (n (1 + \lambda_2^2)))$ and $\sigma^2$ follows an inverse-gamma distribution with shape and scale parameters given by $n / 2$ and
\[
 \frac{n}{2}\left(1 + \frac{\mu_2^2 \lambda_2^2}{\lambda_2^2 + 1}\right),
\]
respectively. The posterior mean and variance of $\beta_2$ are thus: $\mu_2 \lambda_2^2 / (1 + \lambda_2^2)$ and
\[
 \frac{1}{1 + \lambda_2^2} \frac{\left(1 + \frac{\mu_2^2 \lambda_2^2}{\lambda_2^2 + 1}\right)}{n - 2},
\]
respectively.

\textbf{Student distribution.} The log prior density is such that (if we forget about the constants):
\[
  \log g_2\left(\frac{\lambda_2 n^{1/2}}{\ee^{\nu}}(\beta_2 - \mu_2)\right) = -\frac{\gamma + 1}{2} \log\left(1 + \frac{\lambda_2^2 n(\beta_2 - \mu_2)^2}{\gamma \ee^{2\nu}}\right),
\]
where $\gamma$ is the degrees of freedom. The gradient is such that:
\[
 \frac{\partial}{\partial \beta_2} \log g_2\left(\frac{\lambda_2 n^{1/2}}{\ee^{\nu}}(\beta_2 - \mu_2)\right) = -(\gamma + 1) \frac{\lambda_2^2 n (\beta_2 - \mu_2)}{\gamma \ee^{2 \nu} + \lambda_2^2 n (\beta_2 - \mu_2)^2},
\]
\[
 \frac{\partial}{\partial \nu} \log g_2\left(\frac{\lambda_2 n^{1/2}}{\ee^{\nu}}(\beta_2 - \mu_2)\right) = (\gamma + 1) \frac{\lambda_2^2 n (\beta_2 - \mu_2)^2}{\gamma \ee^{2 \nu} + \lambda_2^2 n (\beta_2 - \mu_2)^2}.
\]

\textbf{LPTN distribution.} The log prior density is such that (if we forget about the constants):
\[
  \log g_2\left(\frac{\lambda_2 n^{1/2}}{\ee^{\nu}}(\beta_2 - \mu_2)\right) = \left\{
                                                    \begin{array}{lcc}
                                                      -\frac{\lambda_2^2 n}{2 \ee^{2\nu}}(\beta_2 - \mu_2)^2  & \text{ if } & \frac{\lambda_2 n^{1/2}}{\ee^{\nu}}\abs{\beta_2 - \mu_2}\leq \tau, \\
                                                       -\tau^2 / 2 + \log \tau - \log \abs{\frac{\lambda_2 n^{1/2}}{\ee^{\nu}}(\beta_2 - \mu_2)} & \cr
                                                       \quad + \theta \log \log \tau  - \theta \log \log \abs{\frac{\lambda_2 n^{1/2}}{\ee^{\nu}}(\beta_2 - \mu_2)} & \text{ if } &\frac{\lambda_2 n^{1/2}}{\ee^{\nu}}\abs{\beta_2 - \mu_2}>\tau. \\
                                                    \end{array}
\right.
\]
The gradient is such that:
\[
 \frac{\partial}{\partial \beta_2} \log g_2\left(\frac{\lambda_2 n^{1/2}}{\ee^{\nu}}(\beta_2 - \mu_2)\right) = \left\{
                                                    \begin{array}{lcc}
                                                      -\frac{\lambda_2^2 n}{\ee^{2\nu}}(\beta_2 - \mu_2)  & \text{ if } & \frac{\lambda_2 n^{1/2}}{\ee^{\nu}}\abs{\beta_2 - \mu_2}\leq \tau, \\
                                                      -\frac{1}{\beta_2 - \mu_2} - \theta \frac{1}{\beta_2 - \mu_2} \frac{1}{\log \left|\frac{\lambda_2 n^{1/2}}{\ee^{\nu}}(\beta_2 - \mu_2)\right|} & \text{ if } &\frac{\lambda_2 n^{1/2}}{\ee^{\nu}}\abs{\beta_2 - \mu_2}>\tau, \\
                                                    \end{array}
                                                    \right.
\]
\[
 \frac{\partial}{\partial \nu} \log g_2\left(\frac{\lambda_2 n^{1/2}}{\ee^{\nu}}(\beta_2 - \mu_2)\right) = \left\{
                                                    \begin{array}{lcc}
                                                      \frac{\lambda_2^2 n}{\ee^{2\nu}}(\beta_2 - \mu_2)^2  & \text{ if } & \frac{\lambda_2 n^{1/2}}{\ee^{\nu}}\abs{\beta_2 - \mu_2}\leq \tau, \\
                                                      1 + \theta  \frac{1}{\log \left|\frac{\lambda_2 n^{1/2}}{\ee^{\nu}}(\beta_2 - \mu_2)\right|}  & \text{ if } &\frac{\lambda_2 n^{1/2}}{\ee^{\nu}}\abs{\beta_2 - \mu_2}>\tau. \\
                                                    \end{array}
\right.
\]

\textbf{CTN distribution.} The log prior density is such that (if we forget about the constants):
\[
  \log g_2\left(\frac{\lambda_2 n^{1/2}}{\ee^{\nu}}(\beta_2 - \mu_2)\right) = \left\{
                                                    \begin{array}{lcc}
                                                      -\frac{\lambda_2^2 n}{2 \ee^{2\nu}}(\beta_2 - \mu_2)^2  & \text{ if } & \frac{\lambda_2 n^{1/2}}{\ee^{\nu}}\abs{\beta_2 - \mu_2}\leq \kappa, \\
                                                       -\kappa^2 / 2 & \text{ if } &\frac{\lambda_2 n^{1/2}}{\ee^{\nu}}\abs{\beta_2 - \mu_2}>\kappa. \\
                                                    \end{array}
\right.
\]
The gradient is such that:
\[
 \frac{\partial}{\partial \beta_2} \log g_2\left(\frac{\lambda_2 n^{1/2}}{\ee^{\nu}}(\beta_2 - \mu_2)\right) = \left\{
                                                    \begin{array}{lcc}
                                                      -\frac{\lambda_2^2 n}{\ee^{2\nu}}(\beta_2 - \mu_2)  & \text{ if } & \frac{\lambda_2 n^{1/2}}{\ee^{\nu}}\abs{\beta_2 - \mu_2}\leq \kappa, \\
                                                      0 & \text{ if } &\frac{\lambda_2 n^{1/2}}{\ee^{\nu}}\abs{\beta_2 - \mu_2}>\kappa, \\
                                                    \end{array}
                                                    \right.
\]
\[
 \frac{\partial}{\partial \nu} \log g_2\left(\frac{\lambda_2 n^{1/2}}{\ee^{\nu}}(\beta_2 - \mu_2)\right) = \left\{
                                                    \begin{array}{lcc}
                                                      \frac{\lambda_2^2 n}{\ee^{2\nu}}(\beta_2 - \mu_2)^2  & \text{ if } & \frac{\lambda_2 n^{1/2}}{\ee^{\nu}}\abs{\beta_2 - \mu_2}\leq \kappa, \\
                                                      0 & \text{ if } &\frac{\lambda_2 n^{1/2}}{\ee^{\nu}}\abs{\beta_2 - \mu_2}>\kappa. \\
                                                    \end{array}
\right.
\]

\end{document}